\documentclass[reprint,amssymb,amsmath,amsfonts,superscriptaddress,aps,30pt]{revtex4-2}
\usepackage[utf8]{inputenc}
\usepackage{mathtools}
\usepackage{graphicx}
\usepackage{grffile}
\usepackage[caption=false]{subfig}
\usepackage{bm}
\usepackage{hyperref}
\usepackage{color} 
\usepackage{soul} 
\usepackage{natbib}
\usepackage{setspace}

\usepackage[export]{adjustbox}
\usepackage{floatrow}

\makeatletter
\newsavebox\myboxA
\newsavebox\myboxB
\newlength\mylenA

\newcommand*\xoverline[2][0.75]{%
    \sbox{\myboxA}{$\m@th#2$}%
    \setbox\myboxB\null
    \ht\myboxB=\ht\myboxA%
    \dp\myboxB=\dp\myboxA%
    \wd\myboxB=#1\wd\myboxA
    \sbox\myboxB{$\m@th\overline{\copy\myboxB}$}
    \setlength\mylenA{\the\wd\myboxA}
    \addtolength\mylenA{-\the\wd\myboxB}%
    \ifdim\wd\myboxB<\wd\myboxA%
       \rlap{\hskip 0.5\mylenA\usebox\myboxB}{\usebox\myboxA}%
    \else
        \hskip -0.5\mylenA\rlap{\usebox\myboxA}{\hskip 0.5\mylenA\usebox\myboxB}%
    \fi}
\makeatother

\begin{document}

\title{From Hi-C Contact Map to Three-dimensional Organization of Interphase Human Chromosomes}

\author{Guang Shi}
\affiliation{Department of Chemistry, University of Texas at Austin, 78712}
\author{D. Thirumalai}
\email{dave.thirumalai@gmail.com}
\affiliation{Department of Chemistry, University of Texas at Austin, 78712}

\begin{abstract}
The probability of two loci, separated by a certain genome length, being in contact can be inferred using the Chromosome Conformation Capture (3C) method and related Hi-C experiments.  How to go from the contact map, a matrix listing the mean contact probabilities between a large number of pairs of loci, to an ensemble of three-dimensional structures is an open problem. A solution to this problem, without assuming an assumed energy function, would be the first step in understanding the way nature has solved the packaging of chromosomes in tight cellular spaces. We created a theory, based on polymer physics characteristics of chromosomes and the maximum entropy principles, referred to as HIPPS (Hi-C-Polymer-Physics-Structures) method, that allows us to calculate the 3D structures solely from Hi-C contact maps. The first step in the HIPPS method is to relate the mean contact probability ($\langle \bar{p}_{ij}\rangle$) between loci $i$ and $j$ and the average spatial distance, $\langle \bar{r}_{ij}\rangle$.  This is a difficult problem to solve because the cell population is heterogeneous, which means that a given contact exists only in a small unknown fraction of cells. Despite the population heterogeneity, we first prove that there is a theoretical lower bound  connecting  $\langle p_{ij}\rangle$ and $\langle \bar{r}_{ij}\rangle$ via a power-law relation. We show, using simulations of a precisely solvable model, that the overall organization is accurately captured by constructing the distance map from the contact map even when if the cell population is highly heterogeneous, thus justifying the use of the lower bound. In the second step, the mean distance matrix, with elements $\langle \bar{r}_{ij}\rangle$s, is used as a constraint in the maximum entropy principle to obtain the joint distribution of spatial positions of the loci. Using the two steps, we created an ensemble of 3D structures for the 23 chromosomes from lymphoblastoid cells using the measured contact maps as inputs.  The HIPPS method  shows that conformations of chromosomes are heterogeneous even in a single cell type. The differences in the conformational heterogeneity of the same chromosome in different cell types (normal as well as cancerous cells) can also be quantitatively discerned using our theory.   We validate the method by showing that the calculated volumes of the 23 chromosomes from the predicted 3D structures are in  good agreement with experimental estimates.  Because the method is general, the 3D structures for any species may be calculated directly from the contact map without the need to assume a specific polymer model, as is customarily done. 
\end{abstract}

\maketitle

\section*{Introduction}
The question of how chromosomes are packed in the tight space of the cell nucleus has taken center stage in genome biology, largely due to the spectacular advances in experimental techniques. In particular, the routine generation of a large number of contact maps, reporting on the probabilities that pairs of loci separated by varying genomic lengths are in proximity,  for many species using the remarkable Hi-C technique \cite{LiebermanAiden2009,Dixon2012,Sexton2012,Jin2013,Dekker2013,rao20143d} has provided us a glimpse into  the organization of genomes. A high contact count between two loci means that they interact with each other more frequently compared to ones with low contact count. Thus, the Hi-C data describes the chromosome structures in statistical terms expressed in terms of a contact matrix. An element in the contact matrix is the probability ($\langle p_{ij}\rangle$) that two loci $i$ and $j$ (genomic length is $|i-j|$) is in contact.  The Hi-C data provide only a two-dimensional (2D) representation of the multidimensional organization of the chromosomes. How can we go beyond the genomic contact information to 3D distances between the loci, and eventually the spatial location of each locus is an important unsolved problem. Imaging techniques, such as Fluorescence \textit{In Situ} Hybridization (FISH) and its variations, are the most direct way to measure the spatial distance and coordinates of the genomic loci \cite{wang2016spatial}. But currently, imaging techniques are limited in scope because they only provide information on a small number of loci pairs. In contrast, the Hi-C technique yields average contact probabilities  for a large number of loci pairs.  Is it possible to harness the power of the Hi-C technique to construct, at least approximately, the 3D structures of chromosomes? A major problem with straight forward use of the Hi-C data arises due to cell population heterogeneity  (referred to as PH).  By PH, we mean that a given contact is present in only an (unknown) fraction of cells. This means that there is no straight forward relation connecting  the mean distance ($\langle \bar{r}_{ij}\rangle$) between loci $i$ and $j$ and $\langle p_{ij}\rangle$~\cite{Shi2019}. Because a given contact is not present in all the cells, it also implies that there is conformational heterogeneity (CH) in the chromosome structures. Despite the prevalence of PH, we answer the question posed above in the affirmative by building on the precise results for an exactly solvable Generalized Rouse Model for chromosomes \cite{bryngelson1996internal, Shi2019}, and by using the theoretical distance distribution describing the chromosomes. Unlike many previous studies, we do not assume any energy function to model chromosomes. 

Many data-driven approaches have been developed to reconstruct 3D structures of genomes from Hi-C data \cite{Duan2010,Kalhor2011,Rousseau2011,Zhang2013,Hu2013,Varoquaux2014,Lesne2014,Tjong2016} (see the summary in \cite{Hua2018} for additional related studies). Although these methods are insightful, they do not take the polymer nature of chromosomes into consideration. Therefore, it would be difficult to calculate distance distributions between the loci, measured using imaging experiments, using this approach. On the other hand, polymer models of chromosomes \cite{Giorgetti2014, Zhang2015} usually use Monte Carlo or Molecular Dynamics simulation with an assumed energy function with parameters that have to be calculated (typically) by fitting the simulation results to Hi-C data. In these cases, certain parameters such as bond length and monomer size need to be set arbitrarily to reduce the complexity of the model. Moreover, these studies have not calculated the coordinates of the individual loci in chromosomes using only the Hi-C data as the input. Here, based on analytically solvable generalized rouse model (GRM), we create a method using polymer characteristics of chromosomes and maximum entropy principle to calculate the structures of chromosomes solely from Hi-C data. Recently, in a work \cite{LeTreut2018} that is closely related to certain aspects of the present study, it was assumed that the energy function in GRM (referred to as Gaussian Effective Model in \cite{LeTreut2018}) describes the chromosomes. The spring constants between the loci determined to match the measured contact map. However, we do not assume any energy function, but use characteristics that describe the polymeric properties of  the chromosomes to generate the distance map, which is then used in conjunction with the maximum entropy principle to construct 3D structures from Hi-C data.

Translating the contact map to 3D structures is a difficult problem to solve using solely data-driven approaches without physical considerations that are reflected in the polymeric nature of the chromosomes. One problem is the  difficulty in reconciling Hi-C (contact probabilities) and the FISH data (spatial distances) \cite{Giorgetti2016,fudenberg2017fish,Bickmore2013,williamson2014spatial}. For example, in interpreting the Hi-C contact map, one makes the intuitively plausible assumption that a loci pair with high contact probability must also be spatially close. However, it has been demonstrated using Hi-C and FISH data that high contact frequency does not always imply proximity in space \cite{Giorgetti2016,fudenberg2017fish,Bickmore2013,williamson2014spatial}. Elsewhere \cite{Shi2019}, we showed that because a given contact is present only in certain cells (PH),  a one-to-one relation between contact probability and spatial distance between a pair of loci does not exist. The discordance between Hi-C and FISH experiments makes it difficult to extract the ensemble of 3D structures of chromosomes using Hi-C data alone without taking into account the physics driving the condensed state of genomes. Even if one were to construct polymer models that produce results that are consistent with Hi-C contact maps, certain features of the chromosome structures would be discordant with the  FISH data, reflecting the heterogeneous genome organization\cite{Finn2019}. Thus, one has to contend with two kinds of heterogeneities, which we refer to as population heterogeneity (PH) and conformational heterogeneity (CH).

Despite the difficulties alluded to above, we have created a theory, based on the theoretical distribution of distances for polymers and the principle of maximum entropy to determine the 3D structures solely from the Hi-C data. The resulting physics-based data-driven method, which translates Hi-C data through polymer physics to 3D coordinates of each locus, is referred to as HIPPS (Hi-C-Polymer-Physics-Structures).  
The purposes of creating the HIPPS method are two-fold. (1) We first establish that there is a lower theoretical bound for $\langle \bar{r}_{ij}\rangle$ expressible in terms of a calculable non-linear function involving the contact probability even in the presence of PH. In other words, we prove that $\langle \bar{r}_{ij}\rangle \ge \phi(p_{ij})$ where we compute $\phi(p_{ij})$ using familiar polymer physics concepts.   We establish this relationship using the Generalized Rouse Model for Chromosomes (GRMC) for which accurate simulations can be performed. (2) However, mean spatial distances, $\langle r_{ij}\rangle$s, between a large number of loci pairs do not give the needed 3D structures. In addition, it is important to determine  the variability in chromosome structures because massive conformational heterogeneity (CH) has been noted both in experiments \cite{Stevens2017,Finn2019} and computations \cite{Shi2019}. In order to solve this non-trivial problem, we use the principle of maximum entropy to obtain the ensemble of individual chromosome structures. 

The two-step HIPPS method, which allows us to go from the Hi-C contact map to the three-dimensional coordinates, $\boldsymbol{x}_{i}$ ($i=1,2,3,\cdots,N_{\mathrm{c}}$), where $N_{\mathrm{c}}$ is the length of the chromosome, may be summarized as follows. First, we construct the mean distances $\langle r_{ij}\rangle$s between all loci pairs, $(i,j)$s using a power-law relation connecting $\langle p_{ij}\rangle$s and $\langle r_{ij}\rangle$s.   Then, using the maximum entropy principle, we calculate the distribution $P(\{\boldsymbol{x}_i\})$ with $\langle r_{ij}\rangle$s as constraints, from which  an ensemble of chromosome 3D structures (the 3D coordinates for all the loci) is determined.

The application of our theory to determine the 3D structure of chromosomes from any species is limited only by the experimental resolution of the Hi-C technique. Comparisons with experimental data for the sizes and volumes of chromosomes derived from the calculated 3D structures are made to validate the theory. Our method predicts that the structures of a given chromosome within a single cell  and in different cell types are conformationally heterogeneous. Remarkably, the HIPPS method can detect the differences in the extent of CH of a specific chromosome between normal and cancer cells. 

\section*{Results}

\textbf{Inferring the mean distance matrix ($\bar{\mathbf{R}}$) from the contact probability matrix ($\mathbf{P}$) for a homogeneous cell population}: The elements, $\bar{r}_{ij}$, of the $\bar{\mathbf{R}}$ matrix give the \textit{mean} spatial distance between loci $i$ and $j$. Note that $r_{ij}$ is the distance value  for one realization of the genome conformation in a homogeneous population of cells. Here, we use homogeneous implies that a given contact is present with non-zero probability in the entire cell population. The elements $p_{ij}$ of the $\mathbf{P}$ matrix is the contact probability between loci $i$ and $j$. We first establish a power-law relation between $\bar{r}_{ij}$ and $p_{ij}$ in a precisely solvable model. For the Generalized Rouse Model for chromosomes (GRMC), described in Appendix A,  the relation between $\bar{r}_{ij}$ and $p_{ij}$ is given by, 

\begin{figure*}[!htb]
\centering
\includegraphics[width=\linewidth]{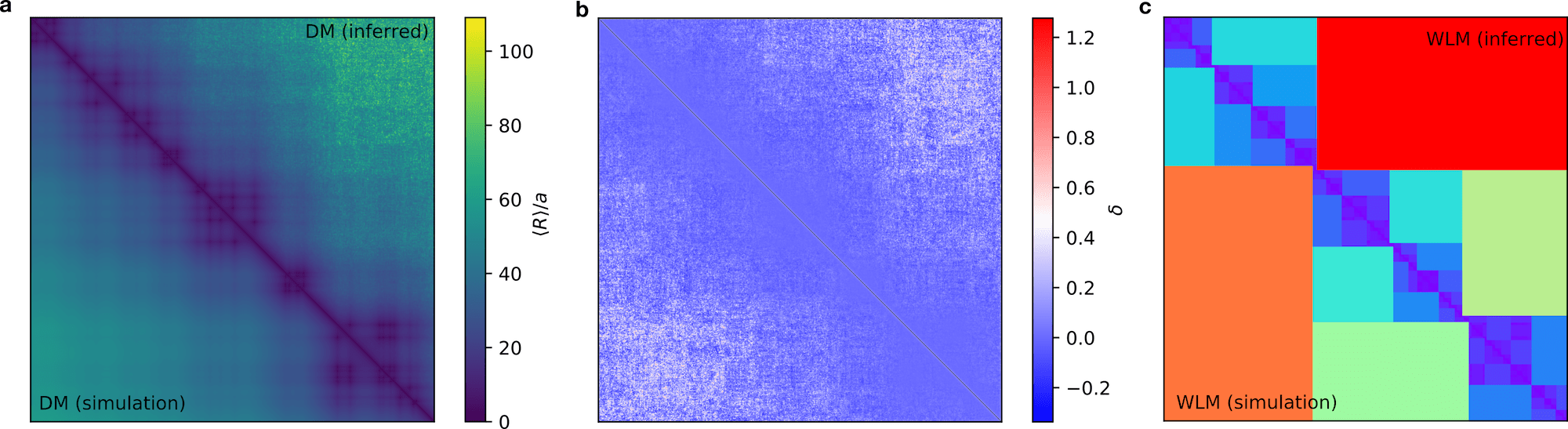}
\caption{Comparison of the distance matrices (DM or $\bar{\mathbf{R}}$) for the GRMC. \textbf{(a)} The simulated $\bar{\mathbf{R}}$ (lower triangle) and the constructed $\bar{\mathbf{R}}$ (upper triangle) are compared side by side. The color bar indicates the value of the mean spatial distance, $\langle R_{mn}\rangle$. The constructed $\bar{\mathbf{R}}$ is obtained by solving Eq.\ref{eq:3} using the contact probability $\mathbf{P}$ (calculated using Eq.\ref{eq:11}). The matrix size is $2000\times 2000$ after the block averaging is applied to the raw data (Appendix C). The threshold value for contact is $r_{c}=2.0a$. The location of the loop anchors are derived from experimental data \cite{rao20143d} over the range from 146 Mbps to 158 Mbps for Chromosome 5 in the Human GM12878 cell line. \textbf{(b)} Relative error $\delta$ is represented as a heatmap. The relative error is calculated as, $\delta = (d_{\mathrm{I}} - d_{\mathrm{S}})/d_{\mathrm{S}}$, where $d_{\mathrm{I}}$ and $d_{\mathrm{S}}$ are the inferred and simulated distances, respectively; $\delta$ increases for loci with large genomic distance indicating the tendency to overestimate the distances for loci pais with small probabilities. \textbf{(c)} Ward Linkage Matrices (WLMs) from the simulation and theoretical predictions, shown in the lower and upper triangle, respectively, are in excellent agreement with each other.}
\label{fig:fig1}
\end{figure*}

\begin{equation}\label{eq:1}
\begin{aligned}
p_{ij}&=\mathrm{erf}(2r_{c}/\sqrt{\pi}\bar{r}_{ij}) - (4\pi/r_{c}\bar{r}_{ij})e^{-4 r_{c}^{2}/\pi \bar{r}_{ij}^{2}}\\
&\equiv f_{\mathrm{GRMC}}(\bar{r}_{ij}).
\end{aligned}
\end{equation}
\noindent where $\mathrm{erf}(\cdot)$ is the error function, and $r_{c}$ is the threshold distance for determining if contact is established. This equation provides a way to calculate the distance matrix ($\bar{\mathbf{R}}$) directly from the contact matrix ($\mathbf{P}$) by inverting $f_{\mathrm{GRMC}}(\bar{r}_{ij})$. Note that $\mathbf{P}$ is inferred only approximately from Hi-C experiments. However, there are uncertainties, in determining both $r_{c}$ due to systematic errors, and $p_{ij}$ due to inadequate sampling, thus restricting the use of Eq.\ref{eq:1} in practice. In light of these considerations, we address the following questions: (a) How accurately can one solve the inverse problem of going from the $\mathbf{P}$ to the $\bar{\mathbf{R}}$? (b) Does the inferred $\bar{\mathbf{R}}$ faithfully reproduce the topology of the spatial organization of chromosomes?  We first answer these questions using the GRMC. 

To answer these two questions, we use a 12 Mbps length segment of Chromosome 5 (146 Mbps to 158 Mbps) as an example. The loop anchors within this segment are derived from the experiment data \cite{rao20143d}. We choose the length of polymer to be 10,000, with each monomer representing 1200 bps. We first constructed the distance map by solving Eq.\ref{eq:1} for $\bar{r}_{ij}$ for every pair $(i,j)$ with contact probability $p_{ij}$. The $\mathbf{P}$ matrix is calculated using simulations of the GRMC, as described in Appendix B. For such a large polymer, some contacts are almost never formed even in long simulations, resulting in $p_{ij}\approx 0$ for some loci pairs. This would erroneously suggest that $\bar{r}_{ij}\to\infty$, as a solution to Eq.\ref{eq:1}. Indeed, this situation arises often in the Hi-C experimental contact maps where $p_{ij}\approx 0$ for many $(i,j)$ pairs. To overcome the practical problem of dealing with $p_{ij}\approx 0$ for several pairs, we apply the block average (a coarse-graining procedure) to $\mathbf{P}$ (described in Appendix C), which decreases the size of the $\mathbf{P}$. This procedure overcomes the problem of having to deal with vanishingly small values of $p_{ij}$ while simultaneously preserving the information needed to solve the inverse problem using Eq.\ref{eq:1}. 

The simulated and constructed distance maps are shown in the lower and upper triangle, respectively in Fig.\ref{fig:fig1}a. We surmise from Fig.\ref{fig:fig1}a that the two distance maps are in excellent agreement with each other. There is a degree of uncertainty for the loci pairs with large mean spatial distance (elements far away from the diagonal (Fig.\ref{fig:fig1}a,b) due to the unavoidable noise in the contact probability matrix $\mathbf{P}$.  The Spearman correlation coefficient between the simulated and theoretically constructed maps is 0.97, which shows that the distance matrix can be accurately constructed. However, a single correlation coefficient is not sufficient to capture the topological structure embedded in the distance map. To further assess the global similarity between the $\bar{\mathbf{R}}$ from theory and simulations, we used the Ward Linkage Matrix \cite{lee2017topological} (WLM), which can capture the hierarchy of the 3D structure. We have previously used WLM to compare the structures of interphase chromosomes \cite{Shi2018}. Fig.\ref{fig:fig1}c shows that the constructed $\bar{\mathbf{R}}$ indeed reproduces the hierarchical structural information accurately. These results  show that the matrix $\bar{\mathbf{R}}$, in which the elements represent the mean distance between the loci, can be calculated accurately, as long as the $\mathbf{P}$ is determined unambiguously. As is well known, this is not possible to do in Hi-C experiments, which renders solving the problem of going from $\mathbf{P}$ to $\bar{\mathbf{R}}$, and eventually the precise three-dimensional structure extremely difficult. \newline

\noindent\textbf{A bound for the spatial distance between loci pairs inferred from the contact probabilities: }The results in Fig.\ref{fig:fig1} show that for a homogeneous system (specific contacts are present in all realizations of the polymer), $\bar{\mathbf{R}}$ can be faithfully reconstructed solely from the $\mathbf{P}$. However, the discrepancies between FISH and Hi-C data in several loci pairs \cite{Fudenberg2017} suggest that there is PH, which means that contact between $i$ and $j$ loci is present in only a fraction of the cells. In this case, which one has to contend with in practice \cite{Shi2019,Finn2019}, the one-to-one mapping between the contact probability and the mean 3D distances (as shown by Eq.\ref{eq:1}) does not hold, leading to the paradox \cite{fudenberg2017fish,Giorgetti2016} that a high contact probability does not imply small inter loci spatial distance. 

Due to PH, one cannot determine the mean 3D distance uniquely from the contact probability, which implies that for certain loci the results of Hi-C and FISH must be discordant. Recently, we solved the Hi-C-FISH paradox by calculating the extent of cell population heterogeneity using FISH data and concepts and theoretical  distribution of distances between monomers along polymers. The distribution of subpopulations could be used to reconstruct the Hi-C data. For a mixed population of cells, the contact probability $p_{ij}$ and the mean spatial distance $\langle \bar{r}_{ij}\rangle$ between two loci $m$ and $n$, are given by,

\begin{align}
\langle \bar{r}_{ij}\rangle &= \sum_m^{S} \eta_{m,ij} \bar{r}_{m,ij} \label{eq:2}\\
\langle p_{ij}\rangle &=\sum_m^{S}\eta_{m,ij} p_{m,ij} \label{eq:3}
\end{align}
\noindent where $\bar{r}_{m,ij}$ and $p_{m,ij}$ are the mean spatial distance and contact probability between $i$ and $j$ in $m^{th}$ subpopulation, respectively. In the above equation, $S$ is the total number of distinct subpopulations, and $\eta_{m,ij}$ is the f the subpopulation fraction for $m$.  The $\eta_{m,ij}$ satisfy the constraint $\sum_m^{S}\eta_{m,ij}=1$. Although there exists a one-to-one relation between $p_{m,ij}$ and $\bar{r}_{m,ij}$ in each of the $m^{th}$ subpopulation, it is not possible to determine $\langle p_{ij}\rangle$ solely from $\langle \bar{r}_{ij}\rangle$ without knowing the values of each $\eta_{m,ij}$ and \textit{vice versa}. 

More generally, if we assume that there exists a continuous spectrum of subpopulations, $\langle \bar{r}_{ij} \rangle$ and $\langle p_{ij}\rangle$ can be expressed as, 

\begin{align}
\langle \bar{r}_{ij}\rangle = \int \mathrm{d}\bar{r}_{ij} K(\bar{r}_{ij})\bar{r}_{ij}\label{eq:4}\\
\langle P_{ij}\rangle = \int \mathrm{d}p_{ij}Q(p_{ij})p_{ij}\label{eq:5}
\end{align}
\noindent where $\bar{r}_{ij}$ and $p_{ij}$ are the mean spatial distance and the contact probability associated with a single population, respectively. $K(\bar{r}_{ij})$ and $Q(p_{ij})$ are the probability density distribution of $\bar{r}_{mn}$ and $p_{mn}$ over subpopulations, respectively.

We have shown \cite{Shi2019} that the paradox arises precisely because of the mixing of different subpopulations. The value $\eta_{m,ij}$, $K(\bar{r}_{ij})$ or $Q(p_{ij})$ in Eq. \ref{eq:2}-\ref{eq:5} in principle could be extracted from the distribution of $\langle \bar{r}_{ij}\rangle$, which can be measured using imaging techniques. However, this is usually unavailable or the data are sparse which leads to the question: Despite the lack of knowledge of the composition of the cell populations (quantitative estimate of PH), can we provide an approximate but reasonably accurate relation between $\langle p_{ij}\rangle$ and $\langle \bar{r}_{ij}\rangle$? In other words, rather than answer the question (a) posed in the previous section precisely, as we did for the homogeneous GRMC, we are seeking an approximate solution. The GRMC calculations provide the insights needed to construct the approximate relation connecting the distance and the contact probability matrices. 

\begin{figure*}[!htb]
\includegraphics[width=.9\linewidth]{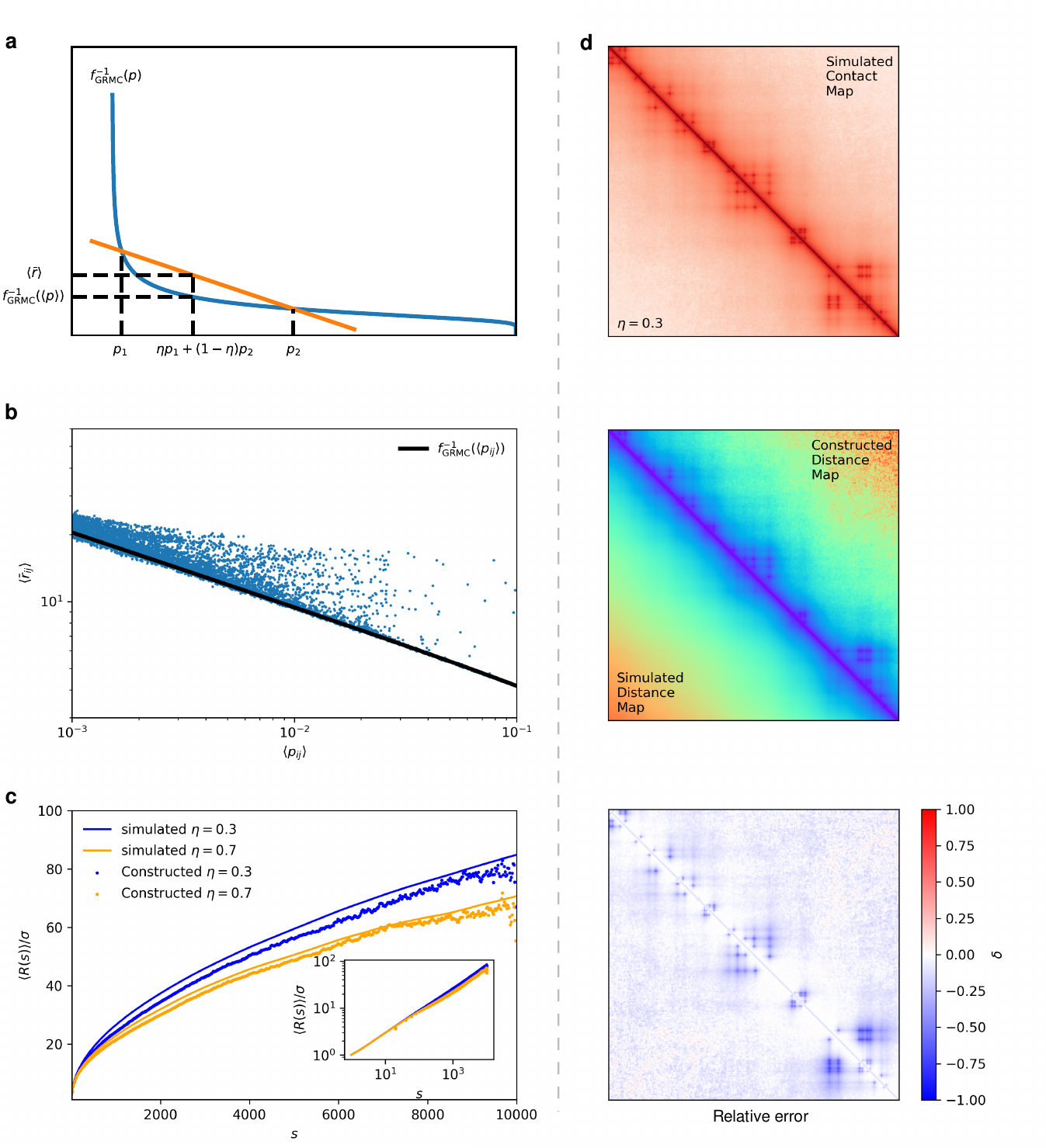}
\caption{(Caption next page.)}
\label{fig:fig2}
\end{figure*}

\textbf{A key inequality:} Let us consider a special case where there are only two distinct discrete subpopulations, and the relation between the $\bar{r}_{ij}(\bar{r})$ and $p_{ij}(p_{ij})$ is given by Eq. \ref{eq:1}. A given contact is present with unity probability in the conformations in one subpopulation and is absent in all the conformations in the other subpopulation. According to Eqs. \ref{eq:2}-\ref{eq:3}, we have $\langle \bar{r}\rangle = \eta \bar{r}_1 + (1-\eta)\bar{r}_2=\eta f_{\mathrm{GRMC}}^{-1}(p_1)+(1-\eta)f_{\mathrm{GRMC}}^{-1}(p_2)$, and $\langle p\rangle = \eta p_1 + (1-\eta)p_2$. Note that $f_{\mathrm{GRMC}}^{-1}$ exists since $f$ is a monotonic function of the argument. Fig.\ref{fig:fig2}a gives a graphical illustration of the inequality $f_{\mathrm{GRMC}}^{-1}(\langle p\rangle) \leq \langle \bar{r}\rangle$. This inequality states that the mean spatial distance of the whole population has a lower bound,  $f_{\mathrm{GRMC}}^{-1}(\langle p\rangle)$, which is the mean spatial distance inferred from the measured contact probability $\langle p\rangle$ as if there is only one homogeneous population (absence of PH). This is a powerful result, which is the theoretical basis for the HIPPS method, allowing us to go from Hi-C data to an ensemble of 3D structures.

The inequality $f_{\mathrm{GRMC}}^{-1}(\langle p\rangle) \leq \langle \bar{r}\rangle$ shows that a theoretical lower bound for $\langle \bar{r}_{ij}\rangle$ exists, given the value of $\langle p_{ij}\rangle$ regardless of the compositions of the whole cell population. The inequality can be generalized to account for arbitrary discrete or continuous distribution of subpopulations. Let us assume that for a homogeneous system, there exists a convex and monotonic decreasing function, $\phi$, relating the contact probability $p$ and the mean spatial distance $\bar{r}$, $\bar{r}=\phi(p)$ (we neglect the suffix $ij$ for better readability). Note that $\phi$ takes the form of Eq. \ref{eq:1} for the GRMC. It can be shown that the following inequality holds (Appendix D),

\begin{align}
\langle \bar{r}\rangle \geq \phi(\langle p\rangle) \label{eq:6}
\end{align}

The above equation (Eq.\ref{eq:6}) shows that the lower bound for the mean spatial distance in the presence of PH is given by the mean spatial distance computed from the measured contact probability as if the cell population is homogeneous. The equality holds exactly only when the population of cells is precisely homogeneous. This finding is remarkably useful in predicting the approximate spatial organization of chromosomes from the Hi-C contact map, as we demonstrate below. Assuming that the single homogeneous population can be described by the GRMC, then the equality in Eq.1 is satisfied. However,  according to Eq. 6, when there are multiple such coexisting populations, the relation $\langle \bar{r}_{ij}\rangle \geq f_{\mathrm{GRMC}}^{-1}(\langle p_{ij}\rangle)$ holds. Thus, the precisely solvable model suggests that the approximate power law relating $\langle p_{ij}\rangle$ and $\langle \bar{r}_{ij}\rangle$ could be used as a starting point in constructing the spatial distance matrices using only the Hi-C contact map for chromosomes.\newline

\noindent\textbf{Validation of the lower bound relating $\langle p_{ij}\rangle$ and $\langle \bar{r}_{ij}\rangle$ in a heterogeneous cell population (PH): } In order to investigate the effect of PH on the quality of the constructed mean distance matrix $\langle \bar{\mathbf{R}}\rangle$ from the contact probability matrix $\langle \mathbf{P}\rangle$, we simulated a model system with two distinct cell populations. One has all the CTCF mediated loops present (with fraction $\eta$), and the other is a polymer chain without any loop constraints (with fraction $1-\eta$) (See Appendix A for simulation details). We used the lower bound, $f_{\mathrm{GRMC}}^{-1}(\langle p_{ij}\rangle )$, to infer $\langle \bar{r}_{ij}\rangle$ from $\langle p_{ij}\rangle$. The results, shown in Figs.\ref{fig:fig2}b,c,d, provide a numerical verification of the theoretical lower bound linking the contact probability and the mean spatial distance. Fig.\ref{fig:fig2}b shows the scatter plot for $\langle \bar{r}_{ij}\rangle$ versus $\langle p_{ij}\rangle$ from the simulation. The theoretical lower bound, $f_{\mathrm{GRMC}}^{-1}(\langle p_{ij}\rangle )$ is shown for comparison. Fig.\ref{fig:fig2}b shows that the lower bound holds. Using the $f_{\mathrm{GRMC}}^{-1}(\langle p_{ij}\rangle)$, we calculated the $\langle \bar{\mathbf{R}}\rangle$  (see Fig.\ref{fig:fig2}d from the simulated $\langle \mathbf{P}\rangle$). Comparison between the inferred and the simulated $\langle \bar{\mathbf{R}}\rangle$ (middle ad bottom in Fig.\ref{fig:fig2}d) shows that the difference between the two $\bar{\mathbf{R}}\rangle$s is large near the loops, resulting in an underestimate of the spatial distances. This occurs because the constructed $\langle \bar{\mathbf{R}}\rangle$ is obtained from the simulated $\langle \mathbf{P}\rangle$, which is sensitive to the PH. The difference matrices show that, although the constructed $\langle \bar{\mathbf{R}}\rangle$  underestimated the spatial distances around the loops, most of the pairwise distances are hardly affected. This exercise for the GRMC  justifies the use of the lower bound as a practical guide to construct $\langle \bar{\mathbf{R}}\rangle$  from the $\langle \mathbf{P}\rangle$.

To show that the constructed $\langle \bar{\mathbf{R}}\rangle$ using the lower bound gives a good global description of the chromosome organization, we also calculated the often-used quantity $\langle R(s)\rangle$, the mean spatial distance as a function of the genomic distance $s$, as an indicator of the average structure (Fig.\ref{fig:fig2}c). The calculated $\langle R(s)\rangle$ differs only negligibly from the simulation results. Notably, the scaling of $\langle R(s)\rangle$ versus $s$ is not significantly altered (inset in Fig.\ref{fig:fig2}c), strongly suggesting that constructing the $\langle \bar{\mathbf{R}}\rangle$ using the lower bound gives a good estimate of the average size of the chromosome segment.

\addtocounter{figure}{-1}
\begin{figure*}
\caption{\textbf{(a)} Lower Bound for the mean spatial distance $\langle \bar{r}\rangle$ illustrated graphically. The blue curve is the function $f^{-1}_{\mathrm{GRMC}}$ which exists since $f_{\mathrm{GRMC}}$ is a monotonic function. The orange line is the secant line between the points $(p_1, f^{-1}_{\mathrm{GRMC}}(p_1))$ and $(p_2, f^{-1}_{\mathrm{GRMC}}(p_2))$.  All the points between $p_1$ and $p_2$ on the x-axis can be expressed as $\eta p_1+(1-\eta)p_2\equiv \langle p \rangle$ for some value of $\eta\in[0,1]$. The y-axis value corresponds to $\langle p\rangle$ is $\eta f^{-1}_{\mathrm{GRMC}}(p_1)+(1-\eta)f^{-1}_{\mathrm{GRMC}}(p_2)\equiv \langle \bar{r}\rangle$ and $f^{-1}_{\mathrm{GRMC}}(\langle p\rangle)$ for the orange line and blue curve, respectively. Notice that for any values of $p_1$, $p_2$ and $\eta$, the orange line is always above the blue curve, which proves the inequality $f^{-1}_{\mathrm{GRMC}}(\langle p\rangle)\leq \langle \bar{r}\rangle$. From the graph, it can also be noted the equality holds only when $p_1=p_2$.  \textbf{(b)} Scatter plot for mean pair-wise spatial distances versus the contact probabilities for $\eta=0.3$. Solid black line is the theoretical lower bound, given by the solution $f_{\mathrm{GRMC}}^{-1}(\langle p_{ij}\rangle)$. \textbf{(c)} Plots of $\langle R(s)\rangle$ as a function of the genomic distance, $s$, for $\eta=0.3$ and $0.7$. The inset shows the same data on a log-log scale; $\langle R(s)\rangle$ is calculated using $\langle R(s)\rangle=(1/TM)\sum_{a=1}^{M}\sum_{t=1}^{T}\big(r_{ij}^{(a)}(t)\delta(s-|i-j|)/(N-s)\big)$. The theoretical predictions are in excellent agreement with simulations. \textbf{(d)} Simulated $\langle \mathbf{P}\rangle$ (top), simulated $\langle \bar{\mathbf{R}}\rangle$ and inferred $\langle \bar{\mathbf{R}}\rangle$ side by side (middle), and relative error map (bottom) for $\eta=0.3$ for GRMC. Note that all the maps are block averaged from $N$=10,000 to size $n$=400 as explained in the Appendix C. The inferred $\langle \bar{\mathbf{R}}\rangle$ is obtained using $\langle \bar{r}_{ij}\rangle =f_{\mathrm{GRMC}}^{-1}(\langle p_{ij}\rangle)$. Relative error map is shown with blue color indicating larger error. }
\end{figure*}

\noindent\textbf{Inferring 3D organization of interphase chromosomes from experimental Hi-C contact map: }To apply the insights from the results from the GRMC to determine the 3D structures of chromosomes, we conjecture that a power-law relation \cite{wang2016spatial,Shi2018}, relating the contact probability $\langle p_{ij}\rangle$ and the spatial distance$\langle \bar{r}_{ij}\rangle$, holds generally for chromosomes. Thus, we write, 

\begin{equation}\label{eq:7}
\langle \bar{r}_{ij}\rangle = \Lambda \langle p_{ij}\rangle^{-1/\alpha}
\end{equation}
\noindent where the coefficients $\alpha$ and $\Lambda$ are unknown.  Again, note that the $\langle \cdot\rangle$ and $\bar{\cdot}$ represent the average over subpopulations and the average over individual conformations in a single subpopulation, respectively. In a homogeneous system, the equalities $\langle \bar{r}\rangle=\bar{r}$ and $\langle p\rangle=p$ hold. For the GRMC, $\Lambda=r_{c}$ and $\alpha=3.0$.  For a self-avoiding polymer, $\alpha\approx 3.71$ for two interior loci that are in contact (see Appendix E). Based on experiments \cite{wang2016spatial} and simulations using the Chromosome Copolymer Model  \cite{Shi2018} a tentative suggestion could be made for a numerical value for $\alpha\approx 4.0$. Given the paucity of data needed to determine $\alpha$, we follow the experimental lead \cite{wang2016spatial} and set it to 4.0.  We show below that the power-law relation given in Eq.\ref{eq:7} provides a way to infer the approximate 3D organization of chromosomes from the experimental Hi-C contact map.\newline

\begin{figure}
\includegraphics[width=\linewidth]{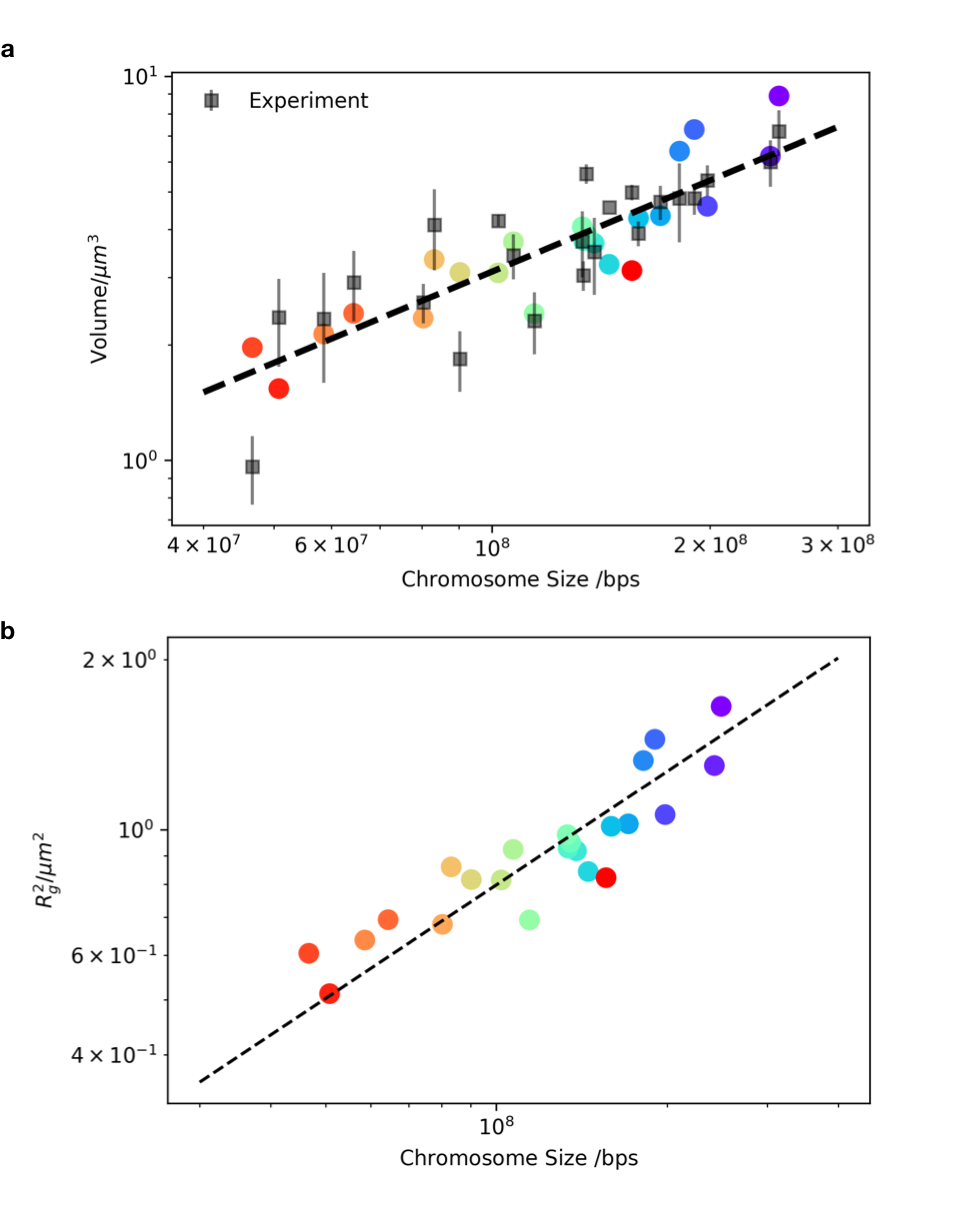}
\caption{\textbf{(a)} Plot of the square of the radius of gyration $R_{g}^{2}$ as a function of the chromosome size. The dashed line is a fit to the data with the slope 0.54, which implies that $R_g\sim N^{0.27}$. The data are for the 23 chromosomes. \textbf{(b)} Volume of each chromosome versus the length in units of base pairs. The experimental values (black squares) are computed using the data in \cite{branco2006intermingling}. The dashed line is the fit to the experimental data with slope=0.8. Volume of each chromosome is calculated using $\lambda V_{\mathrm{nuc}}$ where $\lambda$ is the percentage of volume of the nucleus, $V_{\mathrm{nuc}}$. The values of $\lambda$ are provided in Fig.S5 in \cite{branco2006intermingling}, and $V_{\mathrm{nuc}}=(4/3)\pi r_{\mathrm{nuc}}^{3}$ where $r_{\mathrm{nuc}}=3.5\mathrm{\mu m}$ is the radius of Human lymphocyte cell nucleus \cite{branco2006intermingling}. Volumes of the Chromosomes obtained using theory and computation are calculated using $(4/3)\pi R_{g}^{3}$ (color circles). The Pearson correlation coefficient between  predicted values, without any adjustable parameters, and the experimental data is 0.79.}
\label{fig:fig3}
\end{figure}

\begin{figure*}
\includegraphics[width=0.85\linewidth]{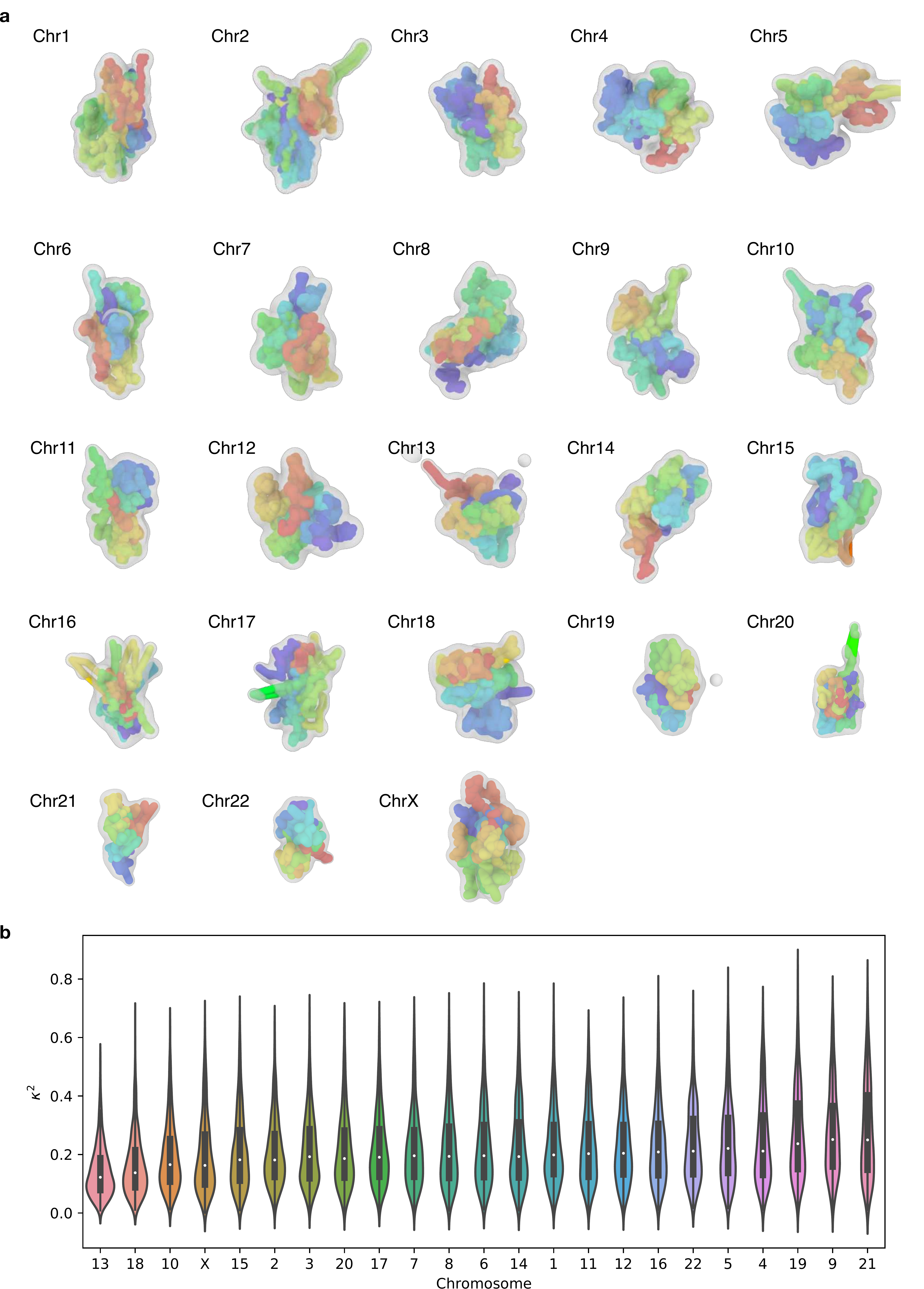}
\caption{\textbf{(a)} Representative 3D reconstructed structures for all the 23 Human interphase chromosomes using the inferred distance matrices, which are calculated using Eq.\ref{eq:7} with $\Lambda=117\mathrm{\ nm}$ and $\alpha=4.0$. The colors encode the genomic position of the loci. The resolution of loci is 100 kbps. Red and purple represent the 5' and 3' ends, respectively. The structures with radii of gyration that are  close to the population average are selected. The structures are rendered using bond radius,  $\Lambda=117$nm. More individual conformations are shown in Fig. \ref{fig:gm12878_individual_conformations}. \textbf{(b)} Violin plot for the relative shape anisotropy $\kappa^2$ (Appendix H) for all the 23 chromosomes. The chromosomes are ordered with increasing of $\langle \kappa^2\rangle$.}
\label{fig:fig5}
\end{figure*}

\noindent{\bf Experimental Validation of Eq\ref{eq:7} and choice of $\alpha$:} Before describing the 3D structures, we first  show that Eq.\ref{eq:7} with $\alpha=4$ is reasonable. To do so  we calculated the square of the radius of gyration of all the 23 chromosomes using $R_{g}^{2}=(1/2N_{c}^{2})\sum_{i,j}\langle \bar{r}_{ij}\rangle^{2}$. The dashed line in Fig.\ref{fig:fig3}a is a fit of $R_{g}^{2}$ as a function of chromosome size, which yields $R_{g}\sim N_{c}^{0.27}$ where $N_{c}$ is the length of the chromosome. For a collapsed polymer, $R_{g}\sim N_{c}^{1/3}$ and for an ideal polymer to be $R_{g}\sim N_{c}^{1/2}$. The exponent $0.27\lesssim 1/3$ suggests that chromosomes adopt highly compact, space-filling structures, which is also vividly illustrated in Fig.\ref{fig:fig5}. To ascertain if the unusual value of 0.27 is reasonable, we computed the volume of each chromosome using $(4/3)\pi R_{g}^{3}$ and compared the results with experimental data \cite{branco2006intermingling}. The scaling of chromosome volumes versus $N_{c}$ calculated from the predicted 3D chromosome structures is in excellent agreement with the experimental data (Fig.\ref{fig:fig3}b). 

Since the value of $\Lambda$ (Eq.\ref{eq:7}) is unknown, we estimate it by minimizing the error between the calculated chromosome volumes and experimental measurements. We find that $\Lambda=117\mathrm{\ nm}$, which is the approximate size of a locus of 100 kbps (the resolution of the Hi-C map used in the analysis). It is noteworthy that the genome density computed using the value of $\Lambda = (100\cdot10^{3}/(4/3)\pi \Lambda^{3})\mathrm{bps}\cdot\mathrm{nm}^{-3}=0.015\mathrm{bps}\cdot\mathrm{nm}^{-3}$ is consistent with the typical average genome density of Human cell nucleus $0.012\mathrm{bps}\cdot\mathrm{nm}^{-3}$ \cite{Rosa2008}. The value of $\Lambda$ does not change the scaling but only the absolute size of chromosomes.  \newline

\noindent\textbf{Generating ensembles of 3D structures using the maximum entropy principle}: The great variability in the genome organization (CH) has been noted before \cite{Stevens2017,Finn2019, Shi2019}. To determine the structural heterogeneity of the chromosomes, we ask the question: how to generate an ensemble of structures consistent with the mean pairwise spatial distances between the loci? More precisely, what is the joint distribution of the position of the loci, $P(\{\boldsymbol{x}_i\})$, subject to the constraint that the mean pairwise distance is $\langle ||\boldsymbol{x}_i-\boldsymbol{x}_j||\rangle=\langle \bar{r}_{ij} \rangle$? Generally, there exists an infinite number of $P(\{\boldsymbol{x}_i\})$, satisfying the mean pair-wise spatial distance constraints. We seek  the $P^{\mathrm{MaxEnt}}(\{\boldsymbol{x}_i\})$, yielding the maximum entropy among all possible $P(\{\boldsymbol{x}_i\})$s. The maximum entropy principle has been previously used in the context of genome organization \cite{DiPierro2016, Farr2018} for different purposes. We note parenthetically that enforcing the constraints of the mean pairwise distances is equivalent to preservation of the mean squared pairwise distances. In practice, we found that constraining the squared distances, $\langle ||\boldsymbol{x}_i-\boldsymbol{x}_j||^2\rangle=\langle \bar{r}_{ij}^2 \rangle$, yields better numerical convergence. The $P^{\mathrm{MaxEnt}}(\{\boldsymbol{x}_i\})$ subject to the constraints associated with the mean squared pairwise spatial distances is given by,

\begin{equation}\label{eq:8}
    P^{\mathrm{MaxEnt}}(\{\boldsymbol{x}_i\}) =\frac{1}{Z} \mathrm{exp}\big({-\sum_{i<j}k_{ij}||\boldsymbol{x}_i-\boldsymbol{x}_j||^2}\big).
\end{equation}
\noindent In the above equation, $Z$ is a normalization factor, and $k_{ij}$s are the Lagrange multipliers that are chosen so that the average values $\langle||\boldsymbol{x}_i-\boldsymbol{x}_j||^2 \rangle$ match $\langle r_{ij}^2 \rangle$. The latter could  either be inferred from the Hi-C contact map or directly measured in FISH experiments. The merit of the maximum entropy distribution (Eq.\ref{eq:8}) is that it is both data-driven and physically meaningful since the parameters $k_{ij}$ are inferred from experimental data and the term $k_{ij}||\boldsymbol{x}_i - \boldsymbol{x}_j||^2$ may be interpreted  as pair-wise potential energy between two loci $i$ and $j$. Indeed, Eq. \ref{eq:8} is exactly the same as the generalized Rouse model \cite{bryngelson1996internal} where $k_{ij}$s are the spring constants between the genomic loci, which has been used as basis for modeling chromosomes recently \cite{Liu2019}.

The procedure used to generate an ensemble of 3D chromosome structures is the following: First, we compute the mean spatial distance matrix from the contact map using Eq. \ref{eq:7} with $\alpha=4.0$. The value of the scaling factor $\Lambda=117\mathrm{nm}$ was calculated using an additional experimental constraint (see the previous section). Recall that $\Lambda$ only sets the over all length scale but has no effect on the conformational ensemble of the chromosome. Using an iterative scaling algorithm \cite{Darroch1972, berger1997improved}, we obtain the values of $k_{ij}$ (Appendix G). Once the values of $k_{ij}$ are obtained, $P^{\mathrm{MaxEnt}}$ can be directly sampled as a multivariate normal distribution, which can then be used to generate an ensemble of chromosome structures. 

\begin{figure*}[!htb]
\includegraphics[width=\linewidth]{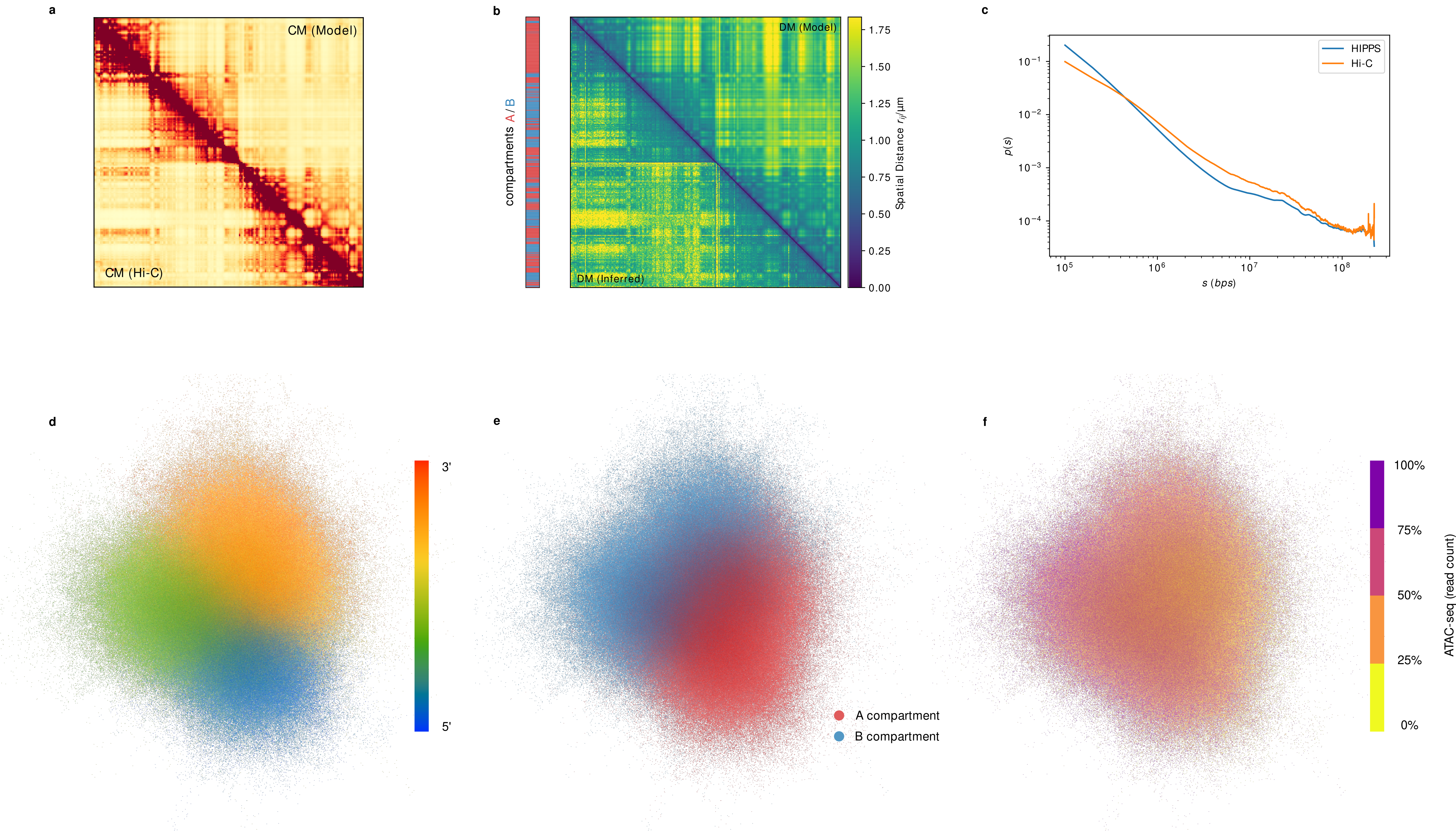}
\caption{\textbf{(a)} Comparison between the Hi-C (lower triangle) and the contact maps calculated from an ensemble of 3D structures for Chromosome 1 using the HIPPS method (($\langle \mathbf{P}\rangle\to \langle \bar{\mathbf{R}}\rangle\to$ 3D structures $\to$ contact map). \textbf{(b)} Comparison between the distance matrix inferred from the Hi-C data (lower triangle) and the distance matrix calculated from an ensemble of 3D structures for Chr1 using the HIPPS method ($\langle \mathbf{P}\rangle\to \langle \bar{\mathbf{R}}\rangle\to$ 3D structures $\to$ mean distance map). A/B compartments, determined using spectral biclustering \cite{Shi2018} are also shown. \textbf{(c)} Comparison between the contact probability profile $P(s)$ inferred from experiment and the calculated curve using the HIPPS method. \textbf{(d)} Superposition of 1,000 3D structures for Chr1.  Each point represents one locus from one conformation. The cloud representation demonstrates the probabilistic nature of chromosome conformation, with color representing the genomic location of the loci along the genome. The resolution of the locus is 100 kbps.  \textbf{(e)} Same cloud point representations as \text{(d)} with colors indicating the A/B compartments. Phase separation between A/B compartments is vividly illustrated. \textbf{(f)} Same as \textbf{(d)} and \textbf{(e)} but with ATAC-seq read counts coded in color}. 
\label{fig:fig6}
\end{figure*}

In Fig.\ref{fig:fig6}a we compare the inferred distance matrix and the distance matrix for Chromosome 1 obtained using the maximum entropy principle. It is visually clear that the two distance matrices are in excellent agreement with each other (see Fig.S2-S7 for the other chromosomes). We should emphasize  that the maximum entropy method described here, in principle, can achieve exact match with the inferred distance matrix. The small discrepancies are due to 1) the quality of convergence, and 2) the intrinsic error in the Hi-C map and the inferred distance matrix derived from it.\newline

\noindent\textbf{Characteristics of the predicted 3D chromosome structures:} To illustrate the applicability of HIPPS, we choose the Hi-C data for cell line GM12878 \cite{rao20143d}. The 3D conformations are specified by  $\boldsymbol{x}_i,i=1,2,3,\cdots,N_c$ where $N_c$ is the number of loci at a given resolution (the centromeres are discarded due to lack to information about them in the Hi-C contact map). The resolution is set to be 100 kbps per monomer. The values of $N_c$ for all the 23 chromosomes are listed in Table.S1. We generated an ensemble of 1,000 structures for each of the 23 Human interphase chromosomes using the HIPPS procedure. Fig.\ref{fig:fig5}a shows the typical conformations for each chromosome. Visually it is clear that there is considerable shape heterogeneity among the chromosomes. To quantify their shapes, we calculated the distribution of relative shape anisotropy $\kappa^2$ (Appendix H). Fig.\ref{fig:fig5}b shows a violin plot for $\kappa^2$ (going from the smallest to the largest value) for the 23 chromosomes.  The chromosomes exhibit considerable variations in $\kappa^2$.  Chromosome 13 is most spherical and chromosome 19, 9 and 21 have the most elongated shape.

{\bf Biological implications based on the 3D structures:} We can draw important conclusions from the calculated 3D structural ensemble for chromosomes with some biological implications that we mention briefly here. 

\textit{Compartments and microphase separation:} The probabilistic representation of the Chromosome 1 structures are shown in Fig.\ref{fig:fig6}b,c,d, where we align all the conformations and superimpose them. First, we note that such  a probabilistic representation demonstrates clear hierarchical folding of chromosomes.  Loci pairs separated by small genomic distance (similar color) are also close in space (Fig.\ref{fig:fig6}b, see Fig.\ref{fig:fig13} for the other chromosomes). Long-range mixing between different loci is avoided, supporting the notion of crumpled globule \cite{grosberg1988role,grosberg1993crumpled,lieberman2009comprehensive}. Second, the chromosome structures exhibit clear microphase separation (different colors are segregated). These are referred to as A and B compartments (Fig.\ref{fig:fig6}c, see Fig.\ref{fig:fig14} for the other chromosomes), representing the two epigenetic states (euchromatin and heterochromatin),  which we  previously determined using the spectral clustering technique \cite{Shi2018}. Each compartment predominantly contains loci belonging to either euchromatin or heterochromatin. Contacts within each compartment are enriched. Interactions between loci within a single epigenetic state  (euchromatin or heterochromatin) are more likely than between loci belonging to distinct epigenetic states.   In the Hi-C data, the compartments appear as a prominent checkerboard pattern in the contact maps.  Fig.\ref{fig:fig6}c shows that the two compartments are spatially separated and organized in a polarized fashion, which is consistent with multiplexed FISH and single-cell Hi-C data\cite{Stevens2017}. 

\textit{Mapping ATAC-seq to 3D structures:} Advances in sequencing technology have been used to infer epigenetic information in chromatin without the benefit of integrating it with structures. In particular, the assay for transposase accessible chromatin using sequencing (ATAC-Seq) \cite{Buenrostro2013} technique provides chromatin accessibility, which in turn provides insights into gene regulation and other functions. The ATAC-seq read counts are obtained and processed (Appendix I) from the data taken from \cite{Buenrostro2013} under GEO accession number GSE47753. Then the data is binned into four quantiles. Fig.\ref{fig:fig6}d shows that the loci with high ATAC and low ATAC signals are spatially segregated. For the majority of the 23 chromosomes, the spatial pattern of ATAC-seq is consistent with the formation of A/B compartments (Fig.\ref{fig:fig15}).  With the structures determined by the HIPPS method  in hand, we mapped the ATAC-Seq data onto an ensemble of conformations for Chromosome 1 from GM 12878 cell in Fig.\ref{fig:fig6}d. It appears that accessibilities in chromosome 1 for various functions (such as nucleosome positioning and transcription factor binding regions) are spatially segregated. Such segregation between loci with high ATAC reads and those with low ATAC reads are also visually clear in other chromosomes as well (Fig.\ref{fig:fig15}). Remarkably, these results, derived from the HIPPS method, follow directly from the Hi-C data {\it without} creating a polymer model with parameters that are fit to the experimental data.

\begin{figure}[!htb]
\includegraphics[width=\linewidth]{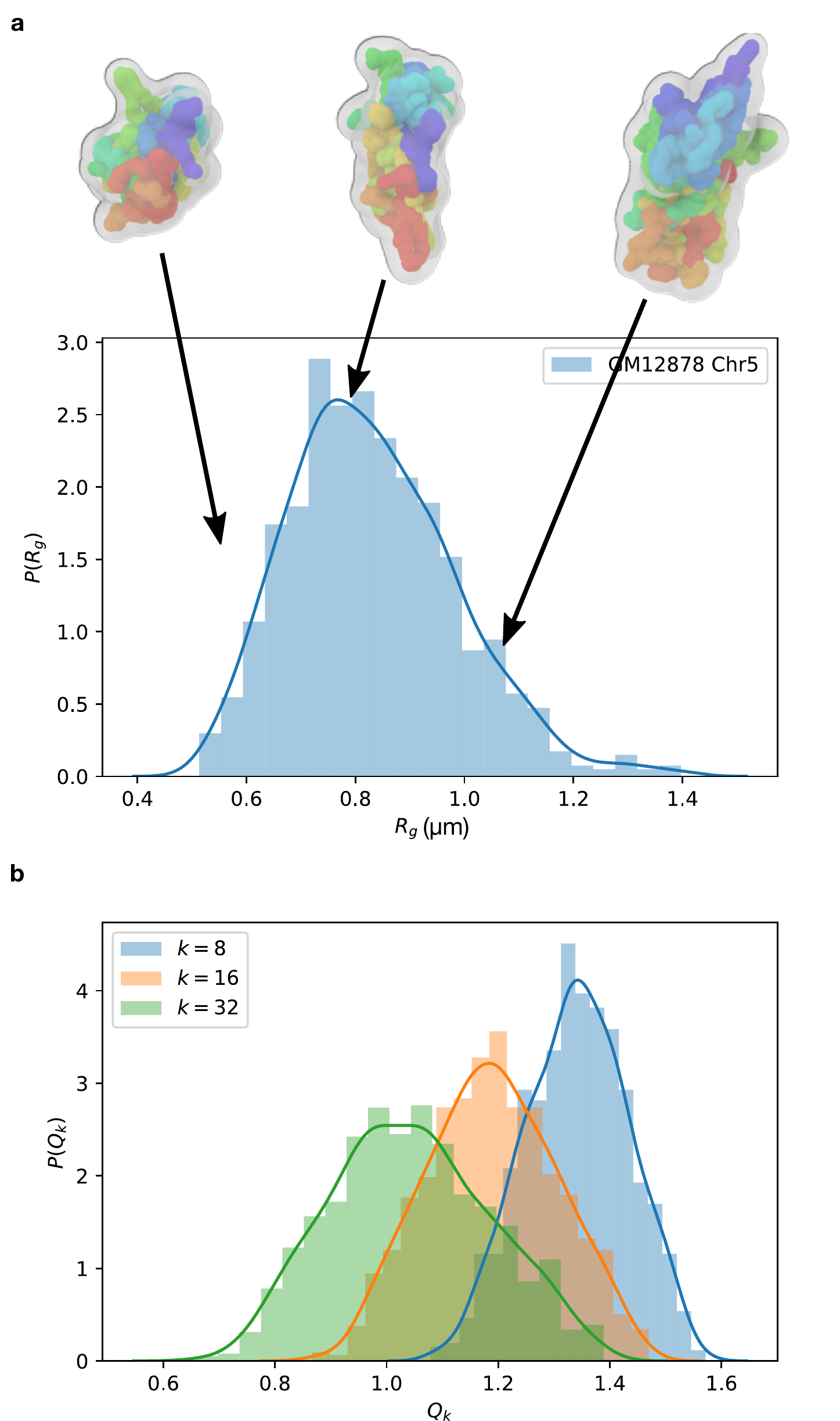}
\caption{\textbf{(a)} Distribution of the radius of gyration, $P(R_g)$, of Chromosome 5 from GM12878 cell type. Three structures whose $R_g$ values are in the 0.15 quantile, 0.5 quantile and 0.75 quantile, respectively are shown. \textbf{(b)} Distribution of the degree of mixing between A/B compartments, $P(Q_k)$ (Eq.\ref{eq:9}), for Chromosome 5.}
\label{fig:fig7}
\end{figure}

{\bf Conformational Heterogeneity (CH) of A/B compartmentalization:} To quantify the extent of CH in chromosomes, we examined the variations among the 1,000 conformations generated for chromosome 5.  Fig.\ref{fig:fig7}a shows the histogram ($P(R_g)$) of $R_g$, the radius of gyration $R_g$.  There is considerable dispersion in $P(R_g)$ in chromosome 5, whose overall shape is anisotropic (see Fig. \ref{fig:fig5}b).  We then wondered what is the degree of variations in the organization of the A/B compartments? Specifically, we are interested in determining whether A/B compartments are spatially separated in a single-cell. To answer this question, we first introduce a quantitative measure of the degree of mixing between A/B compartments, $Q_{k}$,
\begin{equation}\label{eq:9}
    Q_{k} = \frac{1}{N_c}\sum_i \frac{|n_A(i;k)/\hat{n}_A-n_B(i;k)/\hat{n}_B|}{k}
\end{equation}
\noindent where $k$ is the number of the nearest neighbors of loci $i$. In Eq. \ref{eq:9},  $n_A(i;k)$ and $n_B(i;k)$ are the number of neighboring loci belonging to A compartment and B compartment for loci $i$ out of $k$ nearest neighbors, respectively ($n_A(i;k)+n_B(i;k)=k$). With $N_c = (N_A + N_B)$, the fraction of loci in the A compartment is $\hat{n}_A=N_A/N_c$ and $\hat{n}_B=N_B/N_c$ is the fraction in the B compartment where $N_{A}$ and $N_{B}$ are the number of A and B loci, respectively. The $k$ neighbors of $i$ are computed as follows. First, the distance from $i$ to all the loci are calculated. From these distances, the $k$ smallest values are chosen, and this process is repeated for all $i$. Note that $Q_{k}$ is length-scale invariant because it is a function of only the number of nearest neighbors, which allows us to compare the structures with different values of $R_g$ on equal footing.  The value of $Q_{k}=2$ for perfect demixing and $Q_{k}=0$ implies perfect mixing between the A/B compartments. Fig.\ref{fig:fig7}b shows the $P(Q_{k})$ histograms for different values of $k$. The distribution is clearly skewed toward large values, indicating the demixing of the A and B compartments on the population level. However, the distributions also show that a small fraction of single-cell chromosomes conformations with $Q_{k} \approx 0.8$, implying  mixing between A and B compartments to some extent. 
\newline

\noindent\textbf{Chromosome organizations in different cell types}: Since chromosome conformations in a single cell exhibit extensive variations, it is natural to wonder how conformational heterogeneous a given chromosome is in different cells types, and if the HIPPS method can quantify these differences at the single-cell level? We are searching for differences in the conformational heterogeneity of a specific chromosome in different cell types. It is difficult to answer the question posed above precisely because the conformational heterogeneity of a chromosome in a given cell type could overwhelm the analysis.  Furthermore, one has to contend with high-dimensional data (each conformation has 3N coordinates) in the ensemble of conformations. 

\begin{figure}[!htb]
\includegraphics[width=\linewidth]{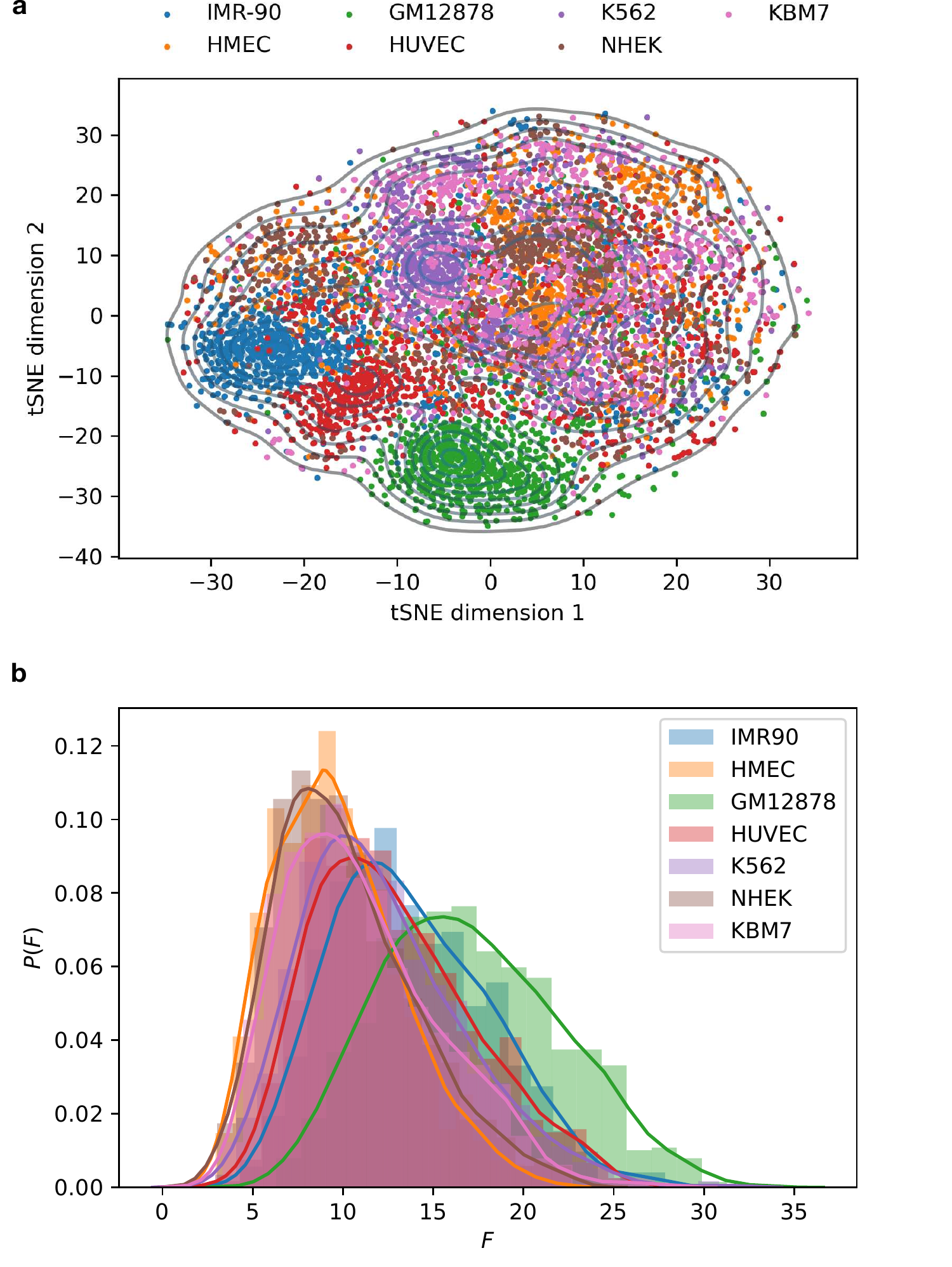}
\caption{\textbf{(a)} tSNE plot for the ensemble of chromosomes 21 structures for 7 cell types (IMR-90, HMEC, GM12878, HUVEC, K562, NHEK, KBM7). We used 1,000 independent conformations for each cell type. A conformation is represented by the distance matrix. The metric used to compare two single chromosomes is the squared Euclidean norm between the distance matrices. \textbf{(b)} The distribution of $F(k)$ (Eq.\ref{eq:10}) for different cell types. We take $k=8$, corresponding to 8 nearest neighbors.}
\label{fig:fig8}
\end{figure}

In order to delineate the differences in the conformational heterogeneities of a specific chromosome in different cell types, we used a machine learning method for analyzing large data \cite{maaten2008visualizing}.  To compare two chromosome conformations, we first normalized the distance matrix such that $\sum_{i,j} r_{ij}^2=1$. By so doing, we eliminate the effect of the overall size of the individual chromosome conformation, thus allowing us to compare them solely in terms of their 3D structures. We generated 1,000 structures for chromosome 21 from 7 cell types using Hi-C data \cite{rao20143d}. Fig.\ref{fig:fig8}a shows the tSNE (t-Distributed Stochastic Neighbor Embedding) plot \cite{maaten2008visualizing} for 7,000 individual chromosome conformations from 7 different cell types (1,000 conformations for each cell type). In Fig.\ref{fig:fig8}a the conformations of chromosome 21 in the 2D tSNE representation are shown as blue (IMR-90), red (HUVEC), and green (GM12878) dots. It is clear that the structural ensembles of chromosome 21 from different cell types have different degrees of overlap with each other. IMR-90 (fibroblast), HUVEC (umbilical vein endothelium),  and GM12878 (lymphoblastoid), which are normal human cells, form compact, distinct clusters with negligible overlap with each other.  In sharp contrast, the conformations of the same chromosome in HMEC (breast epithelial cell), K562 (myeloid leukemia cell in bone marrow), NHEK (epidermal keratinocytes - type of skin cell), and KBM7 (a different leukemia cell) cells display very  large variations.  They are not as compact and their phase space structure in terms of the low dimensional tSNE coordinates show overlapping regions (Fig.\ref{fig:fig8}a). 

To further distinguish between conformational heterogeneity of a given chromosome in different cell types, we computed the value of $Q(k)$ described above for each chromosome, and $F(k)$, which quantifies the multi-body long-range interactions of the chromosome structure. We define $F(k)$ as,
\begin{equation}\label{eq:10}
    F(k)=\frac{1}{k N_c F_0(k)}\sum_i\sum_{j\in m_i(k)} |j-i|
\end{equation}
\noindent where $k$ is the number of nearest neighbors, and $m_i(k)$ is the set of loci that are $k$ nearest neighbors of locus $i$; $F_0(k)=(1/2)(1+k/2)$ is the value of $F(k)$ for a straight chain. From Eq.\ref{eq:10}, it follows that the presence of long-range interaction increases the value of $F(k)$. It is worth  noting that $F(k)$ can also be viewed as a measure of how well the linear relation along the genome is preserved in the 3D structure. Fig.\ref{fig:fig8}b shows the distributions of $F(k)$ for each cell type. GM12878 cell has the largest enrichment of long-range multi-body clusters whereas NHEK and HMEC cells have the least. However, there is extensive overlap between different cell types, as assessed by $F(k)$. Remarkably, we find that there are substantial variations in the structural ensembles of chromosome 21, and by implication others as well, not only within a single cell but also among single cells belonging to different tissues. From our perspective, it is most interesting that the HIPPS method when combined with machine learning techniques can quantitatively predict such differences.\newline

\section*{Evolution of chromosome structures from mitosis to interphase}
\begin{figure*}[!htb]
\includegraphics[width=\linewidth]{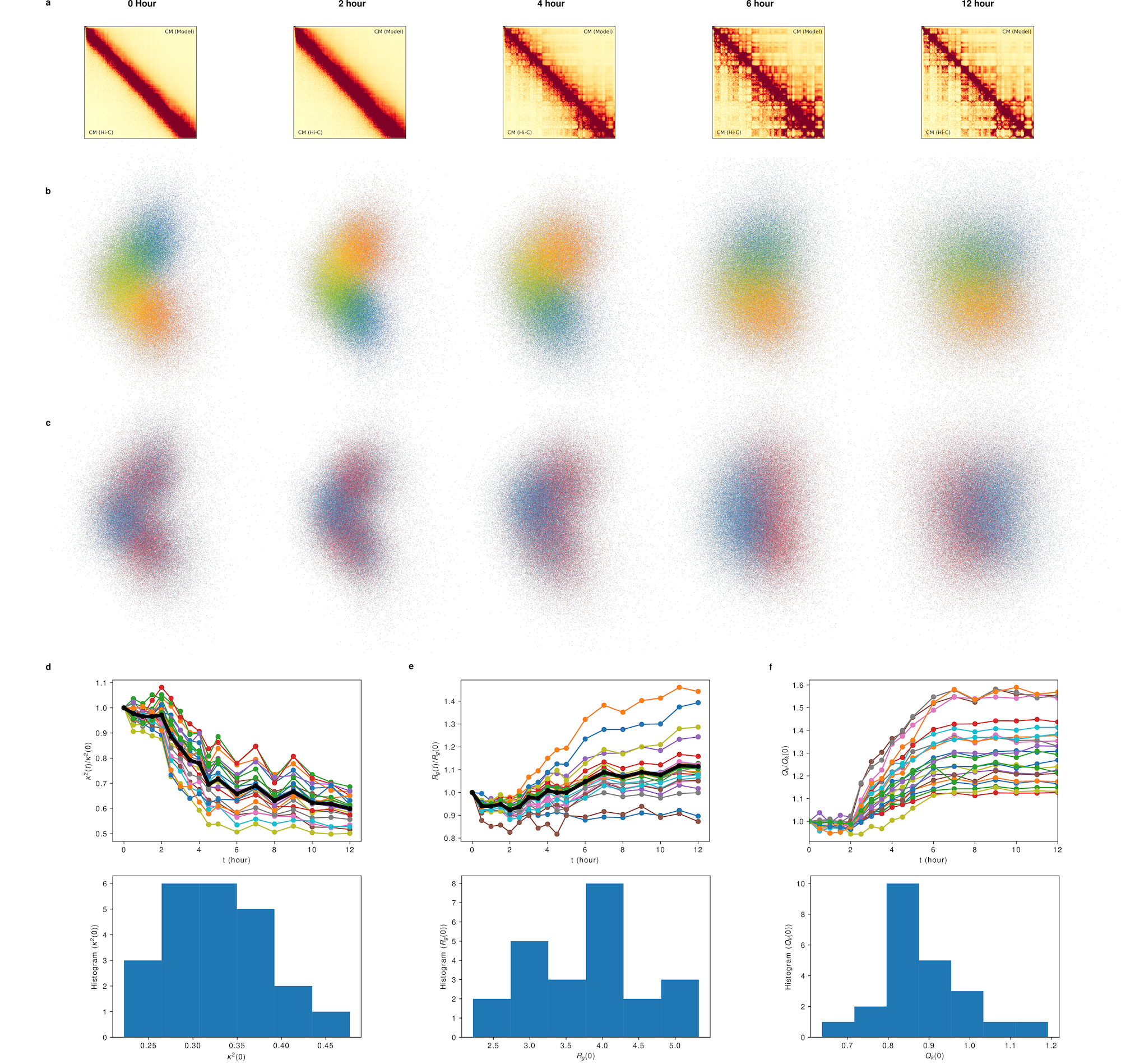}
\caption{\textbf{(a)} Comparison between Hi-C CM and the simulated CM using the HIPPS method for HeLa cell line chromosome 14 at 0, 2, 4, 6, 12 hours after the release from prometaphase. The experiment Hi-C data is taken from Gene Expression Omnibus (GEO) repository under accession number GSE133462. The simulated CM is calculated from an ensemble of 10,000 3D conformations with a chosen contact threshold whose value is determined to minimize the squared difference between the Hi-C and simulated CM. \textbf{(b)} Superposition of 1,000 3D structures for Chr 14 at each time point. Each point represent one locus from one conformation. The color encodes the genomic location of the loci along the genome. \textbf{(c)} Same cloud representation as \textbf{b} with colors indicating the A/B compartments. \textbf{(d)} Top: The change of $\kappa^2$ as a function of time. $\kappa^2$ is normalized by its initial ($t$ = 0) value $\kappa^2(0)$. Bottom: the histogram of $\kappa^{2}(0)$. \textbf{(e)} Top: The change of the radius gyration $R_g$ as a function of time. $R_g$ is normalized by its initial value $R_g(0)$. Bottom: the histogram of $R_{g}(0)$. \textbf{(f)} Top: The time evolution of the degree of compartmentalization $Q_k$, which is calculated using Eq. \ref{eq:9}. Bottom: The histogram of $Q_{k}(0)$.}
\label{fig:abramo2019}
\end{figure*}

We next tested  to ensure that our theory can also be applied to Hi-C data for different time points during the cell cycle. We apply the HIPPS method to the recent Hi-C data from Abramo et al \cite{Abramo2019} in which the Hi-C experiments were performed for HeLa cells at several time points after the arrest of the prometaphase. Fig. \ref{fig:abramo2019}a shows the experiment Hi-C map for HeLa cell chromosome 14  at 6 different time points. The 0-hour corresponds to the arrest of prometaphase. The compartment features emerge during the cell cycle, and are visible after 2 hours. Prior to this time point, the Hi-C contact map is rather featureless. 

Using the HIPPS, we obtained the ensembles of 3D structures corresponding to the 6 time points. Fig. \ref{fig:abramo2019}b shows the superposition of 1,000 3D structures. Similar to Fig.\ref{fig:abramo2019}, each point represent one locus from one conformation. The color encodes the genomic location of each locus along the genome. Individual chromosome conformations are also shown in Fig.\ref{fig:abramo2019_chr14_individual_structures}. Fig. \ref{fig:abramo2019}b shows that the shape of the chromosome changes dramatically during the progression from the mitotic stage to the interphase. At 0 hour, the chromosome adopts a curved cylinder shape while at 12 hours it is  more rounded. To quantitatively investigate the changes in the  chromosome shape and size during the cell cycle, we compute  $\kappa^{2}$ and the radius of gyration $R_{g}$ at various time points for all the chromosomes. The results show that the $\kappa^{2}$ is roughly a constant during the first 2 hours, and slowly decreases as time increases from 2 to 12 hours (Fig. \ref{fig:abramo2019}d).   The size of the  chromosomes (measured by $R_{g}$) , in general, increases after the cell exits mitosis (Fig. \ref{fig:abramo2019}e).

Next we investigate the sequestration  of A/B compartments. As Fig. \ref{fig:abramo2019}a suggests, the compartments are absent during the mitotic and only start appearing after 2 hours. The distribution of A/B locus shown in Fig. \ref{fig:abramo2019}c are largely consistent with the Hi-C data. Visually, the degree of segregation between A/B compartments at 0 hour is less than that at 12 hour end point. To quantify this trend, we compute the $Q_k$ (Eq. \ref{eq:9}) for the available time points for all the chromosomes. We find that $Q_k$ values are nearly constant before 2 hour, and start to increase afterwards and reach a plateau after 6 hours when the segregation between the compartments is complete (Fig. \ref{fig:abramo2019}f).

\section*{Are mitotic chromosomes helical?}

\begin{figure*}[!htb]
\includegraphics[width=\linewidth]{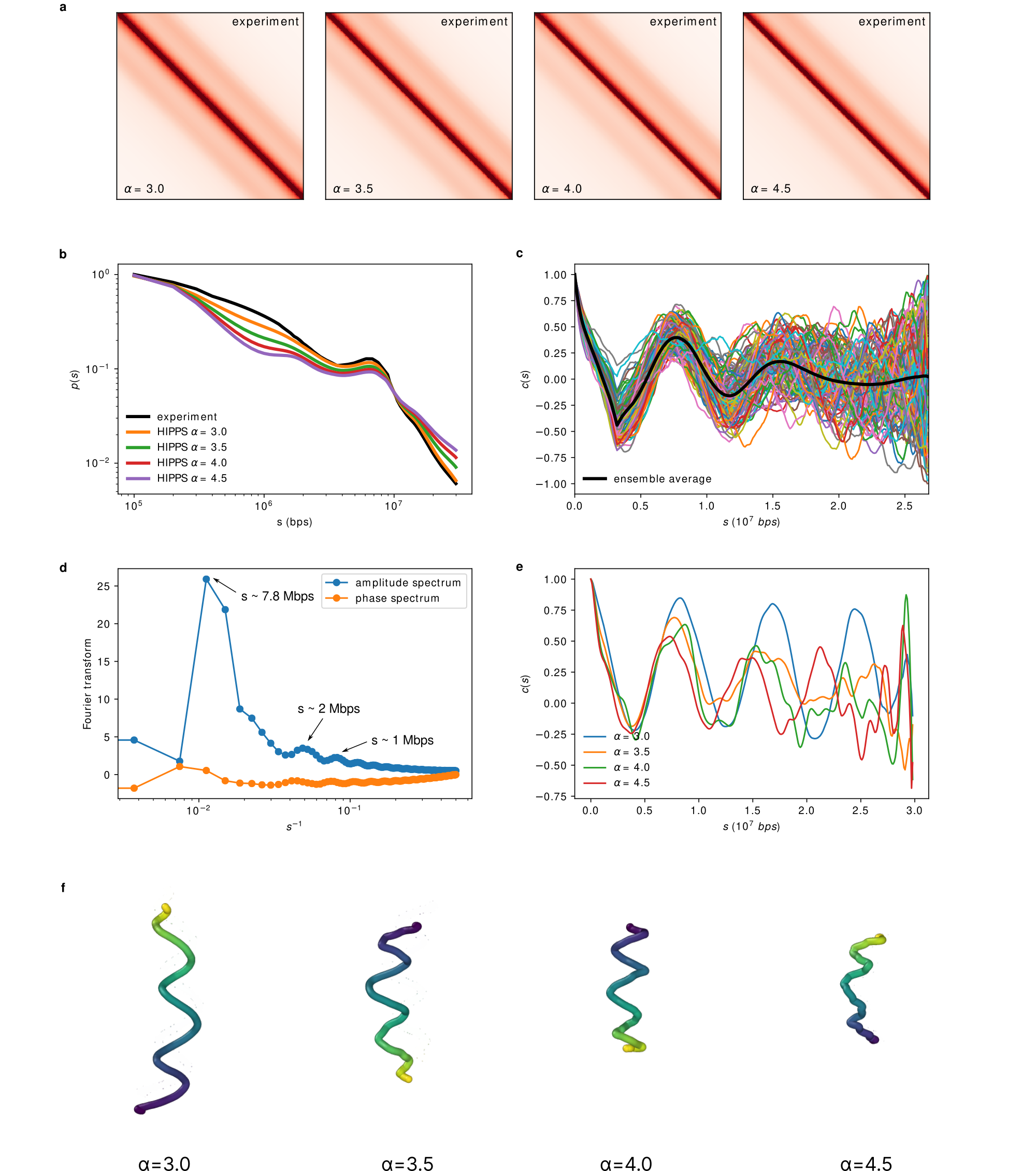}
\caption{\textbf{(a)} Comparison between the theoretical Hi-C CM (described in the text) and the simulated CM for $\alpha=3.0,3.5,4.0,4.5$. \textbf{(b)} Experimental and simulated contact probability profiles $P(s)$ for different values of $\alpha$. \textbf{(c)} Angle correlation function $c(s;d)$ for individual structures and the average curve $\langle c(s;d)\rangle$ (black curve). The value of $d$ is  32. \textbf{(d)} Fourier transform of average $c(s;d)$. Three peaks are marked with corresponding length scales in terms of number base pairs. \textbf{(e)} Angle correlation function $c(s;d)$ with $d=32$ computed from the average structures shown in \textbf{(f)}. \textbf{(f)} The averaged structure for $\alpha=3.0,3.5,4.0,4.5$. A total number of 100,000 random independent individual structures are used to compute the average structure.}
\label{fig:gibcus2018}
\end{figure*}

We have shown that the HIPPS method can be applied to the Hi-C data for different cell states, including the mitosis. We then wondered if the mitotic chromosome structures are helical. Gibcus et al \cite{Gibcus2018} recently suggest that during the prometaphase the Chicken cell chromosomes adopts a helical backbone stabilized by condensin II proteins.  We apply our HIPPS to Gibcus et al data \cite{Gibcus2018} to test if  our HIPPS method can recover such structure. Since the mitotic Hi-C maps are featureless (without any compartments or TADs), we convert the $P(s)$ curve (computed from the Hi-C contact map) to a theoretical Hi-C map to reduce the noise and sampling error in the contact map, and then applied the HIPPS method on the resulting contact map. Furthermore, since the value of $\alpha$ (Eq.\ref{eq:7}) for mitotic chromosomes is not known, we test our model using four different $\alpha$ values, $\alpha=3.0,3.5,4.0,4.5$. The results show that the HIPPS can reasonably reproduce the contact map (Fig. \ref{fig:gibcus2018}a) and the dependence of  $P(s)$ on $s$ (Fig. \ref{fig:gibcus2018}b). For $\alpha=3.0$, the $P(s)$  matches the experimental curve well for all $s$, and quantitatively for $s>10^{6}\ \mathrm{bps}$. The optimal value of $\alpha=3.0$ suggests that mitotic chromosomes may be approximately treated as a near ideal polymer. It is remarkable that without almost no adjustable parameter we can reproduce the experimental $P(s)$ curve including the bump at $s \approx 6\mathrm{Mbps}$ (Fig. \ref{fig:gibcus2018}b). 

To quantitatively investigate whether the mitotic chromosomes structures are helical or have other periodicity, we compute the angle correlation for each individual conformations. The angle correlation is defined as,
\begin{equation}\label{angle_correlation}
c(s,d) = \langle \vec{r}_{i,i+d}\cdot \vec{r}_{i+s,i+s+d}\rangle
\end{equation}

\noindent where $\vec{r}_{i,i+d}$ is the vector between $i^{th}$ and $(i+d)^{th}$ loci, and $d$ is the control parameter. For a perfect helical structure, $c(s,d)$ would exhibit oscillations reflecting the helix pitch as the period.  Fig. \ref{fig:gibcus2018}c shows the results for $c(s,d)$ with $d=32$. The value of $d$ is chosen to be 32 because  the resulting  periodicity is most prominent. Remarkably, we find that there is clear evidence of periodicity. The Fourier transform of $c(s)$ (Fig. \ref{fig:gibcus2018}d) shows that the most prominent peak in the amplitude spectrum is at $s\approx 7.8 \mathrm{Mbps}$ which is in very good agreement with the value reported in Gibcus et al \cite{Gibcus2018}. These authors suggested through a combination of experiments and simulations inspired by the data  that 7-8 Mbps is the length of each helical turn. In addition to this peak, we also find a few less prominent peaks as marked in Fig. \ref{fig:gibcus2018}d, which suggests that the periodicity also are present at $s\approx 2 \mathrm{Mbps}$ and $s\approx 1 \mathrm{Mbps}$.  Finer scale periodicity, which was not reported in Gibcus et al \cite{Gibcus2018}, could be tested using higher resolution experiments.

Next we compute the ``average'' structure defined as follows. First, we generate an ensemble of 100,000 independent individual conformations. Next, we align all structures to a reference structure, with accounting for handedness. Then, the coordinates for each locus in the averaged structure is computed as the mean value of the coordinates of that locus in each individual conformation. The results are shown in Fig. \ref{fig:gibcus2018}f for different value of $\alpha$. Clear helical pattern can be observed for $\alpha=(3.0,3.5)$ whereas it is less transparent for $\alpha=(4.0,4.5)$. Fig. \ref{fig:gibcus2018}e shows the angle correlation $c(s,d)$ with $d=32$ in which the oscillation pattern is clearly observed. We note that such helical pattern is not obvious visually for individual conformation (Fig.\ref{fig:chicken_mitotic_alph30_individual_structure}), suggesting that mitotic chromosome conformations display a degree of heterogeneity with the presence of helical periodicity.

\section*{Discussion and Conclusion}
Using an analytic expression for the distance distribution of distances between monomers in polymers and the principle of maximum entropy, and precise numerical simulations of a non-trivial model, we have provided an approximate solution to the problem of how to construct an ensemble of three-dimensional coordinates of each locus in a chromosome from the measured probabilities ($\langle p_{ij}\rangle$s) that  loci pairs are in contact. The key finding that makes our theory possible is that $\langle p_{ij}\rangle$ is related to $\langle \bar{r}_{ij}\rangle$ through a power law \cite{wang2016spatial,Shi2019}. The inferred mean spatial distances are then used as constraints to obtain an ensemble of structures using the maximum entropy principle. The physically well-tested theory, leading to the HIPPS method, allowed us to use the Hi-C contact map and create an ensemble of three-dimensional chromosome structures without any underlying model. The theory is general enough that sparse data from Hi-C and FISH experiments may be combined to produce the 3D structures of chromosomes for any species. 


The HIPPS method could be improved in at least two ways. First, the theory relies on Eq.\ref{eq:7}, which relates the average contact probability between two loci to the mean distance between them. Even though choosing $\alpha=4.0$ in Eq.\ref{eq:7} provides a reasonable description of the sizes of all the chromosomes it should be treated as a tentative estimate.  More precise data, accompanied by an analytically solvable polymer model containing consecutive loops, as is prevalent in the chromosomes, could produce more accurate structures. Second, as the resolution of Hi-C map improves the size of the contact matrix will not only increase but the matrix would be increasingly sparse because of the intrinsic population and conformational heterogeneities. Thus, mathematical theories for dealing with sparse matrices will have to be utilized in order to extract chromosome structures. 

We should emphasize that if the chromosome structures are used in conjunction with an underlying accurate polymer model then the HIPPS method could also be used to predict structures of chromosomes in single cells, which would shed light on the extent of their conformational heterogeneity. Ultimately, this might well be the single most important utility of our theory.

\section*{Appendix A: Simulation Details}
The GRMC is a variant of a model introduced previously \cite{bryngelson1996internal} as a caricature of physical gels. Recently, we used the GRMC \cite{Shi2019} as the basis to characterize the massive heterogeneity in chromosome organization. The energy function for the GRMC is \cite{Shi2019},
\begin{equation} \label{eq:11}
U(\boldsymbol{r}_{1},...,\boldsymbol{r}_{N})=\sum_{i=1}^{N-1}U_{i}^{S}+\sum_{\{p,q\}}U_{\{p,q\}}^{L}.
\end{equation}
\noindent For the bonded stretch potential, $U_{i}^{S}$, we use,
\begin{equation} \label{eq:12}
U_{i}^{S}=\frac{\kappa}{2}(|\boldsymbol{r}_{i+1} - \boldsymbol{r}_{i}|-a)^{2},
\end{equation}
\noindent where $a$ is the equilibrium bond length. The interaction between the loop anchors is modeled using,
\begin{equation} \label{eq:13}
U_{\{p,q\}}^{L}=\frac{\omega}{2}(|\boldsymbol{r}_{p} - \boldsymbol{r}_{q}| - a)^{2}
\end{equation}
\noindent where the spring constant may be associated with the CTCF facilitated loops. The labels $\{p,q\}$ represent the indices of the loop anchors, which are taken from the Hi-C data \cite{rao20143d}. 

The energy function for the ideal Rouse chain simulated in this work is,
\begin{equation} \label{eq:14}
U(\boldsymbol{r}_{1},...,\boldsymbol{r}_{N})=\sum_{i=1}^{N-1}U_{i}^{S},
\end{equation}
\noindent which is obtained from the energy function for GRMC by eliminating the loop constraints (setting $\omega=0$ in Eq.\ref{eq:13}).

In order to accelerate conformational sampling, we performed Langevin Dynamics simulations at low friction \cite{honeycutt1992nature}. The total number, $N$, of monomers is $10,000$. We simulated each trajectory for $10^{8}$ time steps, and saved the snapshots every $10,000$ time steps. We generated ten independent trajectories, which are sufficient to obtain reliable statistics (see Fig.S8).

\section*{Appendix B: Data analyses of the simulation data}
The contact probability between the $m^{th}$ and $n^{th}$ loci in the simulation is calculated using,

\begin{equation}\label{eq:15}
P_{mn} = \frac{1}{TM}\sum_{a=1}^{M}\sum_{t=1}^{T}\Theta(r_{c}-|\boldsymbol{r}^{(a)}_{m}(t)-\boldsymbol{r}^{(a)}_{n}(t)|),
\end{equation}
\noindent where $\Theta(\cdot)$ is the Heaviside step function, $r_{c}$ is the threshold distance for determining the formation of contacts, the summation is over the snapshots along the trajectory, and $M$ is the total number of independent trajectories, and $T$ is the number of snapshots in a single trajectory. The mean spatial distance between the $i^{th}$ and the $j^{th}$ loci in the simulations is calculated using,

\begin{equation}\label{eq:16}
\langle R_{mn}\rangle = \frac{1}{TM} \sum_{a=1}^{M}\sum_{t=1}^{T}|\boldsymbol{r}^{(a)}_{m}(t) - \boldsymbol{r}^{(a)}_{n}(t)|.
\end{equation}
\noindent The objective is to calculate $\langle R_{mn}\rangle$ from $P_{mn}$ , and to determine, if in so doing, we get reasonably accurate results. Because these quantities can be computed precisely for the GRMC, the $[P_{mn}, \langle R_{mn}\rangle]$ relationship can be rigorously tested.

\section*{Appendix C: Block average}
\begin{figure*}[!htb]
\includegraphics[width=\textwidth]{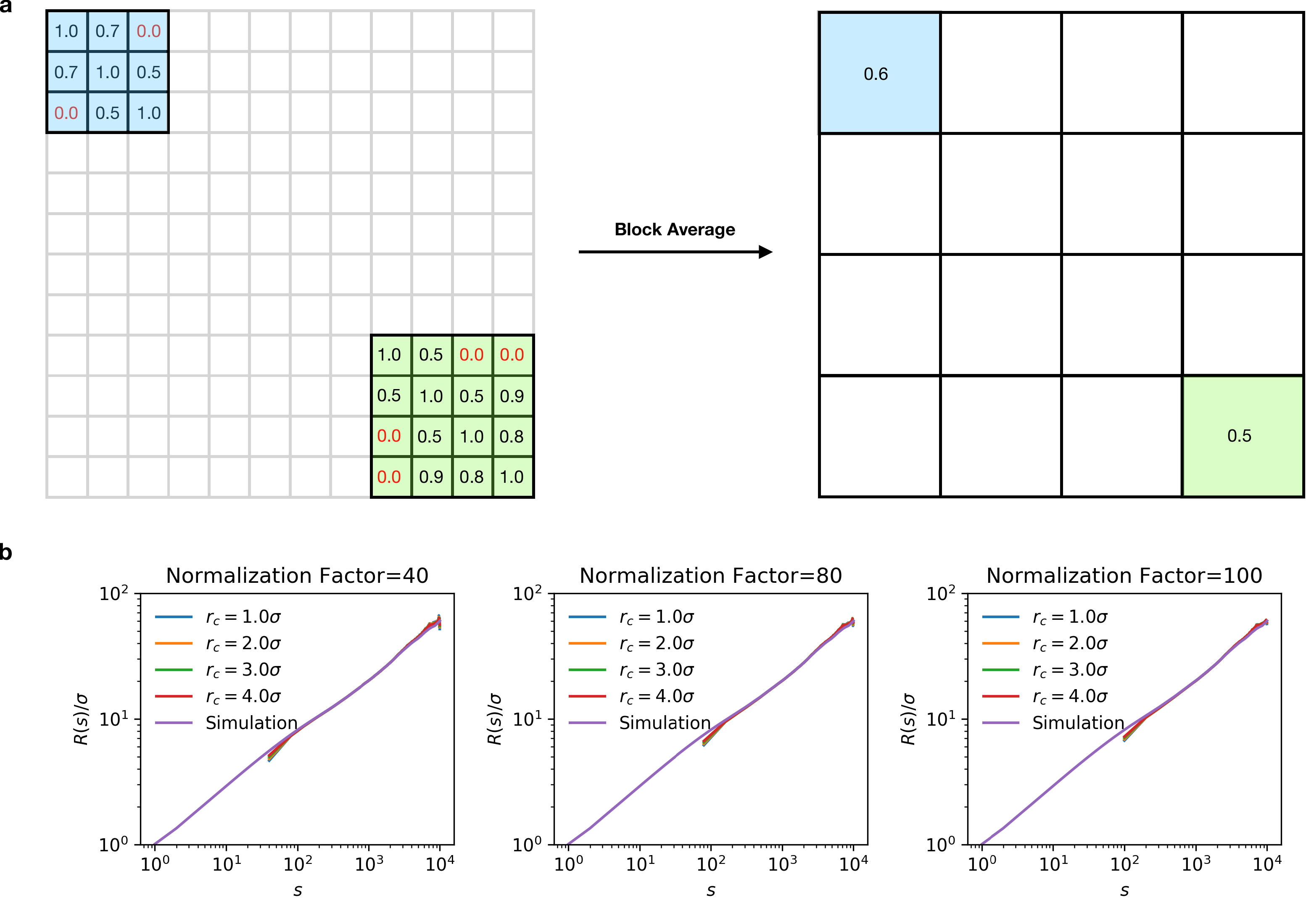}
\caption{\textbf{(a)} Illustration of block average performed on sparse contact map matrix ($\langle \mathbf{P} \rangle$). There are zero value elements in the original $\langle \mathbf{P} \rangle$ (matrix on the left). When constructing the distance matrix, $\langle \bar{\mathbf{R}}\rangle$, from such $\langle \mathbf{P} \rangle$, the zero value contact probability would naively imply that $\langle \bar{r}\rangle\to\infty$. To overcome this problem, we use block averages. The original $N\times N$ $\langle \mathbf{P} \rangle$ are replaced by blocks with size $n$ (red blocks on top left). The value of the matrix element in each block is computed as the mean value of the original elements in each block (matrix on the right). The size of the matrix is reduced from $N$ to $N/n$ where $n$ is the normalization factor. The same procedure could also be applied to $\langle \bar{\mathbf{R}}\rangle$. \textbf{(b)} Block average does not alter the information embedded in the original $\langle \mathbf{P} \rangle$ and the calculated $\langle \bar{\mathbf{R}}\rangle$. $R(s)$ is computed for different values of the normalization factor, $n$. The insensitivity of the results to the block averaging justifies its use in overcoming the problem of missing data points on the $\langle \mathbf{P} \rangle$.}
\label{fig:fig9}
\end{figure*}

Fig.\ref{fig:fig9} shows the procedure used for the block average procedure when dealing with several vanishing (or very small) contact probabilities $P_{mn}$s. Such a method could be used for (almost) any sparse matrix. Let the size of original contact matrix (CM) be $N\times N$. By setting a coarse-grained level $n$, the original CM is divided into blocks, each with size $n\times n$. The new coarse-grained CM is constructed in such a way that  the values of elements in the $(N/n)\times (N/n)$ are the arithmetic average of elements in each block. We then demonstrate that this coarse-graining procedure does not alter the structural information embedded in the original CM.

\section*{Appendix D: Derivation of a lower bound for the spatial distance in terms of contact probability}
Let us use $\bar{\cdot}$ and $\langle \cdot\rangle$ to denote the average over each genome conformations in a single homogeneous population and the average over each individual subpopulations, respectively. The separate averages account for PH and CH.  Here, $\bar{r}_{ij}$ and $p_{ij}$ are the \textit{mean} spatial distance and the contact probability between loci $i$ and $j$ for a single homogeneous (sub)population. $\langle \bar{r}_{ij}\rangle$ and the $\langle p_{ij}\rangle$ are the \textit{mean} spatial distance and the contact probability between loci $i$ and $j$ measured  for the whole population. It is easy to see that if the population is homogeneous, we have $\langle \bar{r}_{ij}\rangle=\bar{r}_{ij}$ and $\langle p_{ij}\rangle = p_{ij}$.

In this appendix, we prove that there exists a theoretical lower bound for $\langle \bar{r}_{ij}\rangle$ for a given value of $\langle p_{ij}\rangle$. We assume that for a homogeneous population, where only one cell population is present, there exists a convex and monotonic decreasing function relating the contact probability between two loci and their mean spatial distance, $\bar{r}_{ij} = \phi(p_{ij})$. For better readability, we will neglect the suffix $ij$ from now on. For a heterogeneous population, the contact probability is calculated as,

\begin{equation}\label{eq:17}
\begin{aligned}
\langle p\rangle &= \int_{0}^{r_c}\int_0^{\infty} \mathrm{d}r \mathrm{d}\bar{r} K(\bar{r})P(r|\bar{r}) \\
&=\int_{0}^{\infty} \mathrm{d} \bar{r} K(\bar{r}) \int_{0}^{r_c} \mathrm{d}r P(r|\bar{r})\\
&=\int_{0}^{1} p K(\phi(p))\frac{\mathrm{d}{\bar{r}}}{\mathrm{d}p}\mathrm{d}p\\
&\equiv \int_{0}^{1} p \psi(p)\mathrm{d}p
\end{aligned}
\end{equation}
\noindent where $K(\bar{r})$ is the distribution of $\bar{r}$ for all the subpopulations (accounts for PH), and $P(r|\bar{r})$ is the distribution of spatial distance for a single subpopulation (accounts for CH) given its mean value $\bar{r}$. $r_c$ is the threshold distance for determining the contact. Note that $p=\int_{0}^{r_c} \mathrm{d}r P(r|\bar{r})$ by definition. $\psi(p)\equiv K(\phi(p))(\mathrm{d} \bar{r}/\mathrm{d}p)$ is the probability measure of $p$ over individual subpopulation. Since $\phi$ is a convex function, according to Jensen's inequality, we have,

\begin{equation}\label{eq:18}
\begin{aligned}
\phi(\langle p\rangle) \leq \langle \phi(p)\rangle = \int \phi(p)\psi(p)\mathrm{d}p
\end{aligned}
\end{equation}

Replace the $\psi(p)$ by $K(\phi(p))(\mathrm{d}\bar{r}/\mathrm{d}p)$. We obtain,
\begin{equation}\label{eq:19}
\begin{aligned}
\phi(\langle p\rangle) &\leq \int \phi(p)K(\phi(p))\frac{\mathrm{d}\bar{r}}{\mathrm{d}p}\mathrm{d}p\\
&=\int \bar{r} K(\bar{r})\mathrm{d}\bar{r}=\langle \bar{r}\rangle
\end{aligned}
\end{equation}
Eq. \ref{eq:19} shows that the lower bound for $\langle \bar{r}\rangle$ is the mean spatial distance inferred from the $\langle p\rangle$ as if the population of genome is homogeneous. In other words there is only one single population without PH.

To demonstrate the validity of Eq. \ref{eq:19}, we consider the special case where there are only two distinct discrete subpopulations. In this case, it is obvious that $\langle \bar{r}\rangle=\eta \bar{r}_1 + (1-\eta)\bar{r}_2$ and $\langle p\rangle=\eta p_1+(1-\eta)p_2$. Note that $\bar{r}_1=\phi(p_1)$ and $\bar{r}_2=\phi(p_2)$. Let us denote $p_1=x$ and $p_2=y$. Given the value of the contact probability $\langle p \rangle$, we show that the lower bound for $\langle \bar{r}\rangle$ is $\phi(\langle p\rangle)$. This is equivalent to the optimization problem,

\begin{equation}\label{eq:20}
\begin{aligned}
&\mathrm{maximize\   }f(x,y)\\
&\mathrm{subject\ to\ }g(x,y)=0
\end{aligned}
\end{equation}
\noindent where $f(x,y)=-\eta \phi(x)-(1-\eta)\phi(y)\equiv - \langle \bar{r}\rangle$ and $g(x,y)=\eta x + (1-\eta) y -\langle p\rangle$. The Lagrange multiplier is $\mathcal{L}(x,y,\phi)=f(x,y)-\phi g(x,y)$. Using the condition that $\nabla_{x,y,\phi}\mathcal{L}(x,y,\phi)=0$, it can be shown that $f(x,y)$ is maximized when $x=y$. Thus, we proved that $\langle \bar{r}\rangle$ is minimized when $p_1=p_2$ and its minimum value is $\phi(\langle p\rangle)$. This is also graphically illustrated in Fig.\ref{fig:fig2}a in the main text.
\newline

\section*{Appendix E: Connection between the contact probability and mean spatial distance}
For a self-avoiding homopolymer,  the distance distribution between two monomers along a polymer chain is \cite{des1980short},
\begin{equation}\label{eq:21}
P(r|\bar{r})=A (r/\bar{r})^{2+g}\mathrm{exp}(-B (r/\bar{r})^{\delta})
\end{equation}
\noindent where $r$ is the distance between two monomers, $\bar{r}$ is the mean distance between them. $g$ is ``correlation hole" exponent, and $\delta$ is related to the Flory exponent by $\delta=1/(1-\nu)$. Given the contact threshold, the contact probability $p$ between the two monomers is
\begin{equation}\label{eq:22}
p=\int_{0}^{r_{c}}P(r|\bar{r})\mathrm{d}r
\end{equation}
\noindent If the contact threshold is small compared to the size of the chain $r\ll \bar{r}$, the integral can be approximately evaluated as,

\begin{equation}\label{eq:23}
\begin{split}
p &= \lim_{r_{c}\to 0}\int_{0}^{r_{c}}P(r|\bar{r})\mathrm{d}r\\
      &= \lim_{r_{c}\to 0}\int_{0}^{r_{c}}A (r/\bar{r})^{2+g}\mathrm{exp}(-B (r/\bar{r})^{\delta})\mathrm{d}r\\
      &\sim \bar{r}^{-(3+g)}
\end{split}
\end{equation}
Thus, the contact probability between two monomers, $p$, is connected to their mean distance $\bar{r}$ by a scaling exponent, $-(3+g)$. For an ideal chain, $g=0$, we recover the asymptotically exact relation $p\sim\bar{r}^{-3}$. For a self-avoiding chain, there are three cases \cite{des1980short}: (i) two monomers are at the two ends of the chain. (ii) one monomer is in the chain interior, while the other is at the end. (iii) two monomers are located in the central part of a chain. The correlation hole exponents corresponding to the three cases \cite{des1980short} are $g_{1}=0.273$, $g_{2}=0.46$ and $g_{3}=0.71$. Thus, we have $p=\bar{r}^{-3.273}$ for the contact between two ends of a self-avoiding chain. $p=\bar{r}^{-3.46}$ for contact between two monomers in case (ii), and $p=\bar{r}^{-3.71}$ for the contacts between two monomer located in the chain interior.

For polymers in poor solvents (likely more relevant to the Human interphase chromosomes), the value of $g$ is not well known. Using simulations, Bohn et al \cite{bohn2009conformational} showed that for an equilibrium collapsed homopolymer chain, $g=-0.11$ for two ends of the chain. This leads to the contact probability between two ends of an equilibrium homopolymer globule and the mean distance $p=\bar{r}^{-2.89}$. But the values of $g$ for scenarios (ii) and (iii) are unknown. In addition, copolymer and out of equilibrium states of chromosomes further complicate the theoretical calculations. Hence, the theoretical estimate of the relation between $p$ and $\bar{r}$ for chromosomes is not known rigorously. Nevertheless, we expect based on the arguments given here that a power law connecting $p$ and $\bar{r}$ ought to exist. We use the relation based on experimental data and our previous study \cite{Shi2018}.
\newline

\section*{Appendix G: Iterative scaling algorithm for maximum entropy principle}

\begin{figure*}
\includegraphics[width=\textwidth]{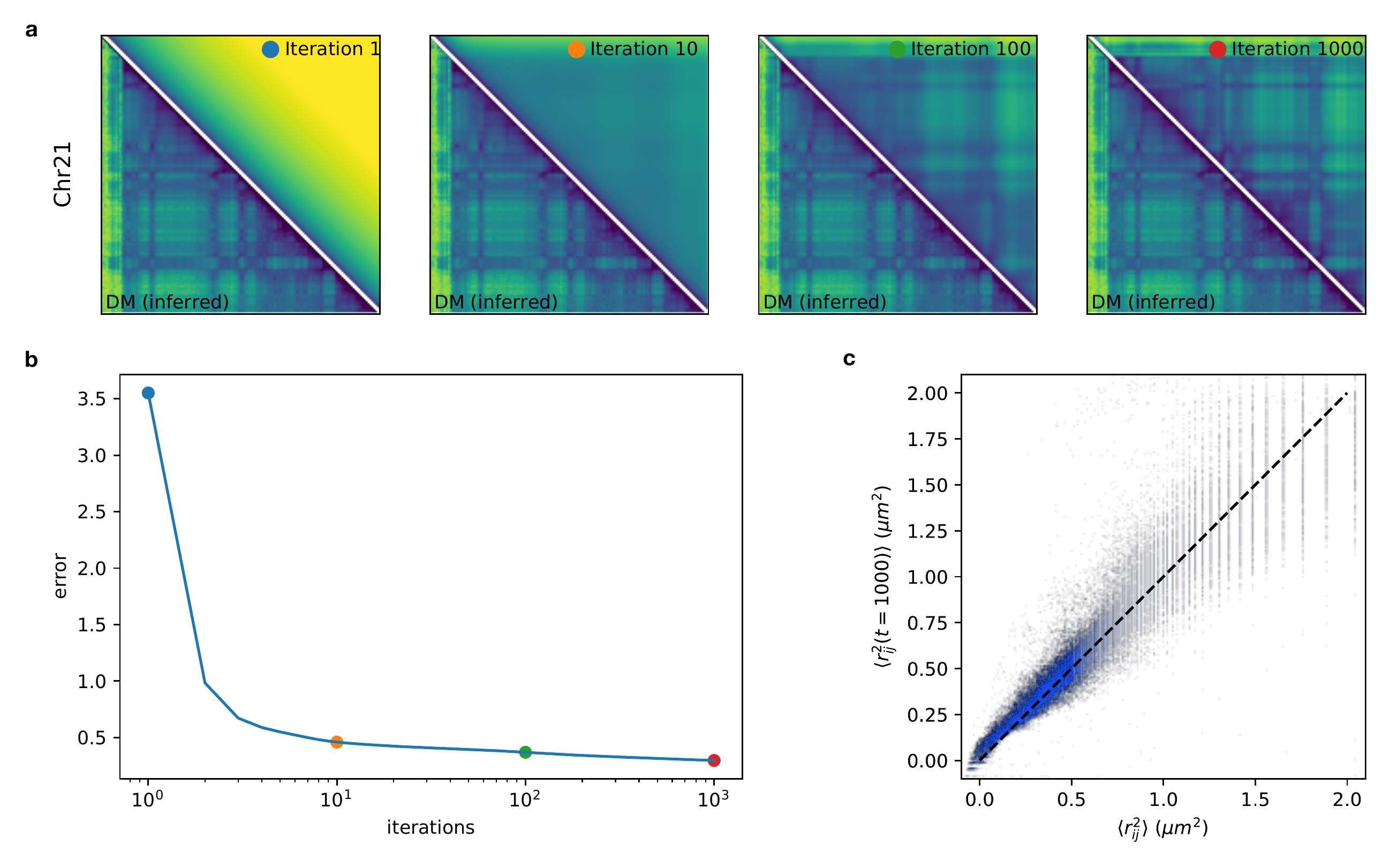}
\caption{\textbf{(a)} Comparison between the targeted distance matrix (lower triangle) and the distance matrix at different iteration steps. At iteration step 1,000, we achieve good agreement with targeted distance matrix. \textbf{(b)} The error as a function iteration steps. The error is defined as the L2 norm between targeted distance matrix and simulated distance matrix. \textbf{(c)} The scatter plot between targeted $\langle r_{ij}^2\rangle$ and $\langle r_{ij}^2(t)\rangle$ at $t=1000$. The pearson correlation coefficient between $\langle r_{ij}^2\rangle$ and $\langle r_{ij}^2(t=1000)\rangle$ is 0.92.}
\label{fig:fig10}
\end{figure*}

Here, we describe the algorithm for obtaining the $k_{ij}$s in Eq.\ref{eq:8}. The algorithm we adopted is iterative scaling \cite{Darroch1972, berger1997improved}. Denote $k_{ij}(t)$ as the value of $k_{ij}$ at $t^{th}$ iteration, it is updated according to,
\begin{equation}\label{eq:30}
k_{ij}(t+1) = k_{ij}(t) + \frac{r}{\sum_{i<j}\langle r_{ij}^2(t)\rangle} \ln \frac{\langle r_{ij}^2(t)\rangle}{\langle r_{ij}^2\rangle}
\end{equation}
\noindent where $r$ is the learning rate. $\langle r_{ij}^2(t)\rangle$ is the average squared pairwise distance at $t^{th}$ iteration and $\langle r_{ij}^2\rangle$ is the targeted squared pairwise distance. Generally, the value of $\langle r_{ij}^2(t)\rangle$ can be estimated by numerical sampling methods, such as Monte-Carlo simulation or Langevin Dynamics, under the values of parameters $k_{ij}(t)$. In this particular case, $\langle r_{ij}^2(t)\rangle$ can be directly computed since $P^{\mathrm{MaxEnt}}$ is a multivariate normal distribution. Following the derivation in our previous work \cite{Shi2019}, 

\begin{equation}
\langle r_{ij}^2(t)\rangle = 3 \sigma_{ij}^2(t)
\end{equation}

\noindent where $\sigma_{ij}^2 = \Omega_{ii}+\Omega_{jj}-2\Omega_{ij}$. $\Omega_{ii}$, $\Omega_{jj}$ and $\Omega_{ij}$ are the elements of the matrix $\bm{\Omega}$ which is defined as $\bm{\Omega}=-\bm{V}\bm{\Lambda}^{-1}\bm{V}^{T}$. $\bm{V}$ and $\bm{\Lambda}$ are computed through the eigendecomposition of the connectivity matrix $\bm{K}$ such that $\bm{K}=\bm{V}\bm{\Lambda}\bm{V}^{T}$. The connectivity matrix $\bm{K}$ is defined as, $K_{ij}=k_{ij}$ for $i\neq j$ and $K_{ii}=-\sum_{j,j\neq i}k_{ij}$.

To demonstrate the effectiveness of the algorithm, Fig.\ref{fig:fig10} shows the comparison between targeted average distance matrix and simulated average distance matrix at different iteration steps. It is clear that after a sufficient number of steps, the simulated distance matrix converges to the targeted one with high accuracy.

\section*{Appendix H: Relative shape anisotropy}
To quantify the shape of each chromosome conformation, we calculate the relative shape anisotropy ($\kappa^2$) uing,
\begin{equation}\label{eq:29}
\kappa^2 = \frac{3}{2}\frac{\lambda_1^2+\lambda_2^2+\lambda_3^2}{(\lambda_1+\lambda_2+\lambda_3)^2} - \frac{1}{2}
\end{equation}
\noindent where $\lambda_{1,2,3}$ are the eigenvalues of the gyration tensor. The bounds for $\kappa^2$ is $0\leq \kappa^2\leq 1$, where $0$ is for highly symmetric conformation and 1 corresponds to a rod.

\section*{Appendix I: Processing ATAC-seq data}
\begin{figure}
\includegraphics[width=\textwidth]{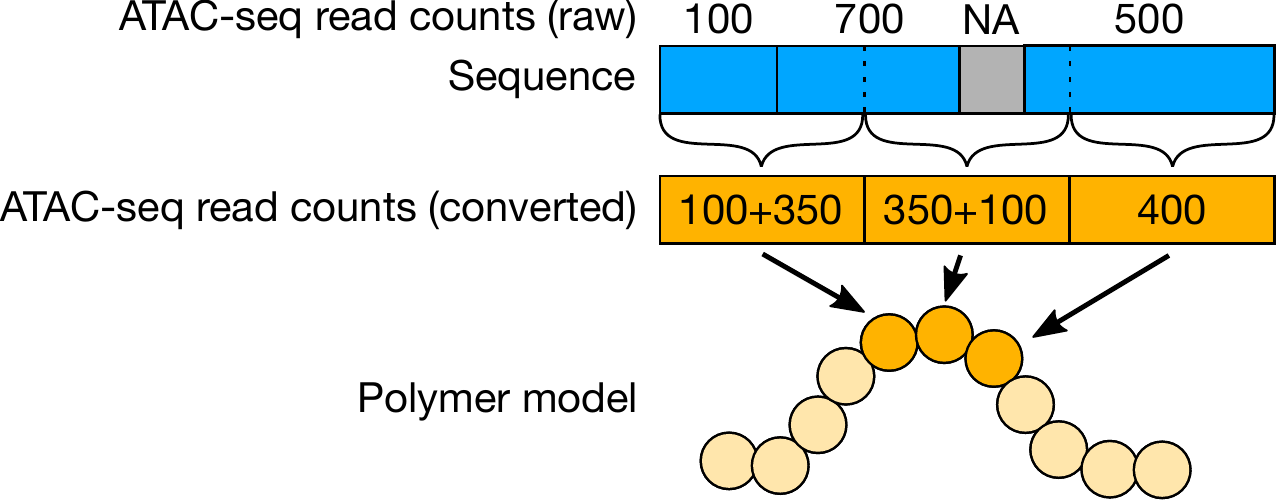}
\caption{The procedure for processing ATAC-seq peak data. The raw ATAC-seq read counts data is illustrated at the top track. Each chromatin segment has a read count value.  The segments are not distributed uniformly, but have different lengths, and have missing parts. In our model, each locus has a fixed genomic length. Thus, to estimate the read counts associated with each locus, we calculate the contribution from the original ATAC-seq segments (blue track) to the segments represented by the locus (yellow track).}
\label{fig:atac-seq-illustration}
\end{figure}

\begin{figure*}
\includegraphics[width=0.8\textwidth]{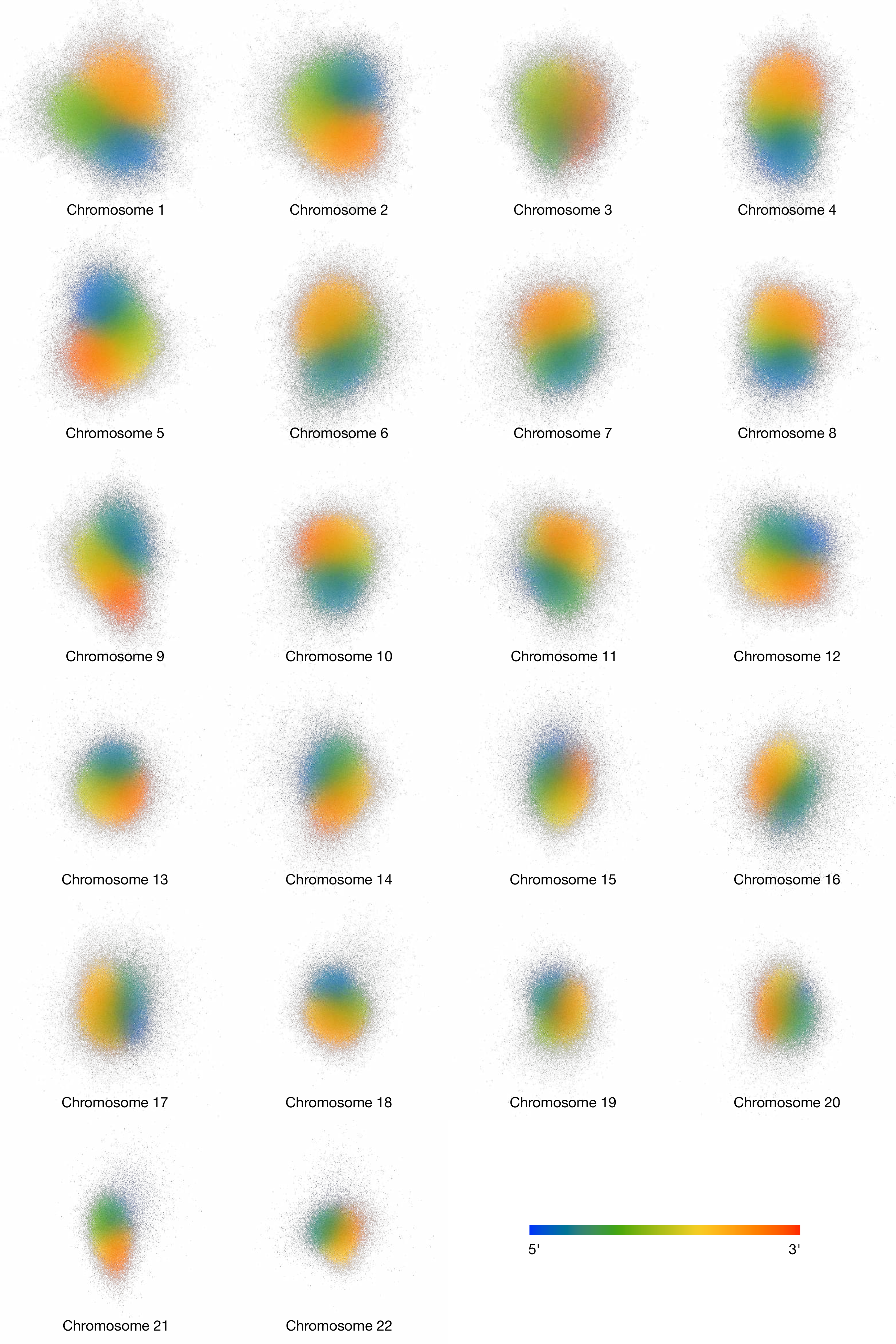}
\caption{Superposition of an ensemble of 3D structures for all 23 chromosomes. A total number of 1,000 conformations are aligned and superimposed for each chromosome. Each point represents one locus from a single conformation, with color representing the genomic location of the locus along the genome}
\label{fig:fig13}
\end{figure*}

\begin{figure*}
\includegraphics[width=0.8\textwidth]{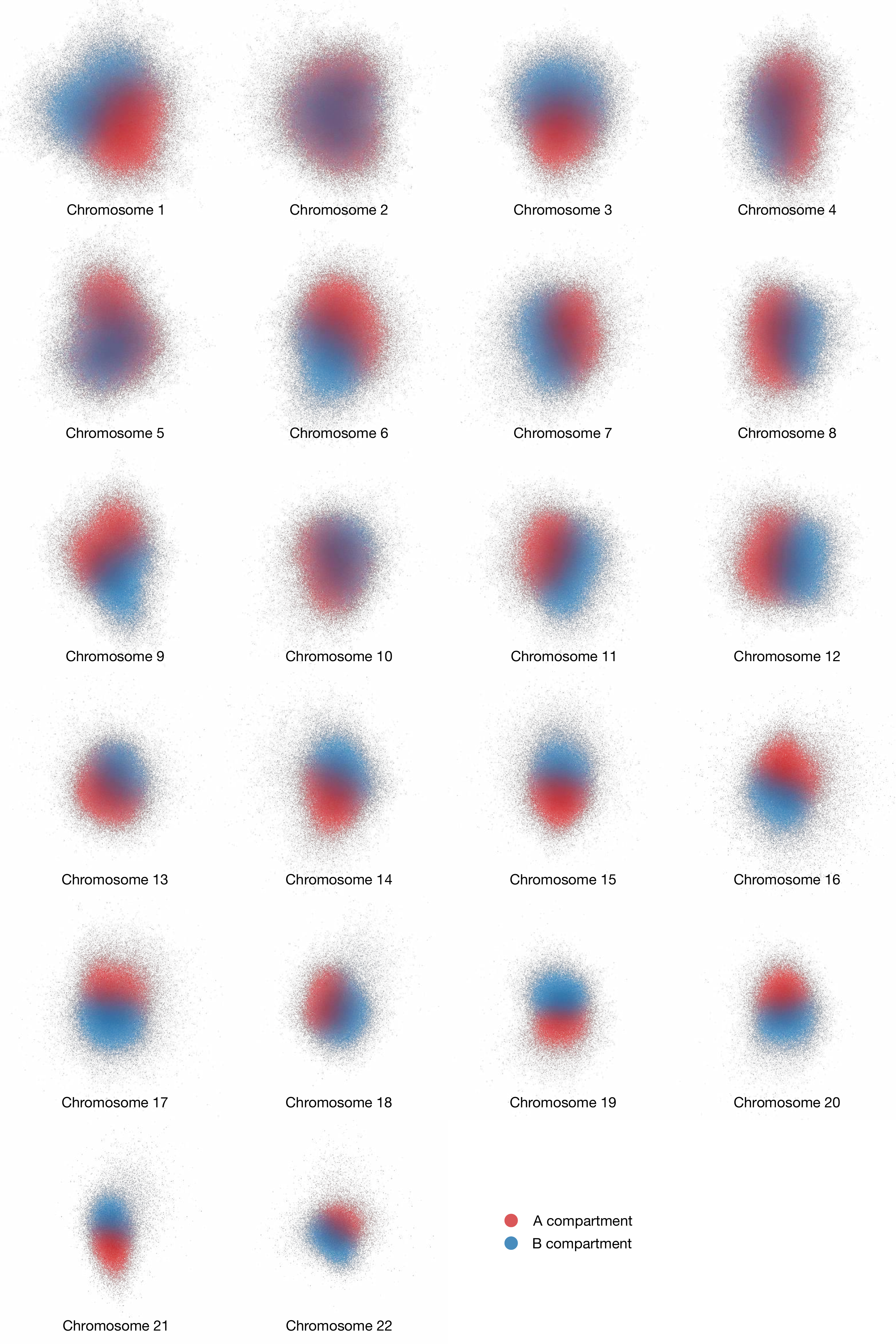}
\caption{Superposition of an ensemble of 3D structures for all 23 chromosomes. A total number of 1,000 conformations are aligned and superimposed for each chromosome. Each point represents a single locus from one conformations, with colors representing the A/B compartments. Note that the A/B compartments do not necessarily correspond to the same epigenetic state across different chromosomes since the assignment of label A or label B is arbitrary.}
\label{fig:fig14}
\end{figure*}

\begin{figure*}
\includegraphics[width=0.8\textwidth]{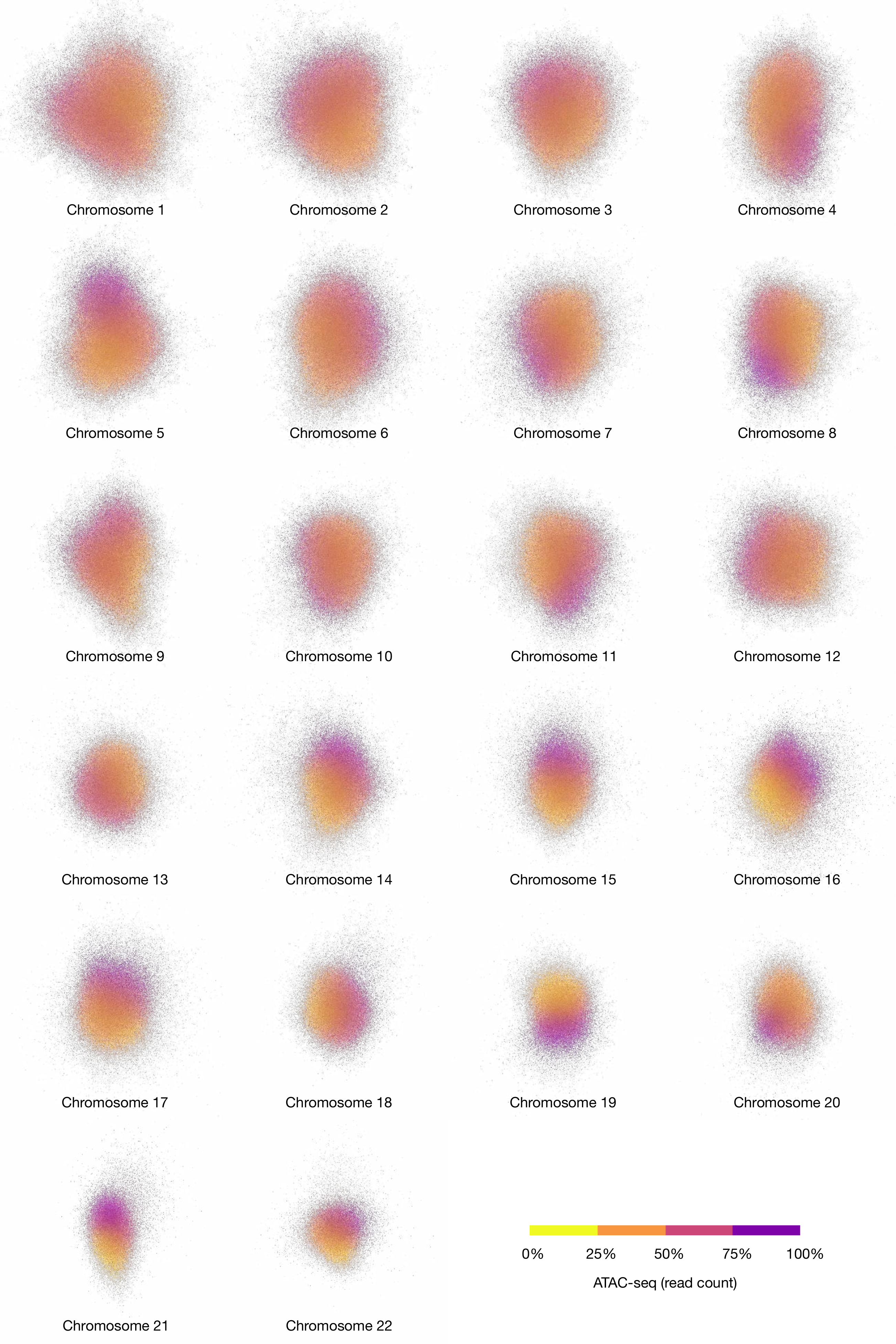}
\caption{An ensemble of 3D structures for all 23 chromosomes obtained from 1,000 conformations that are aligned and superimposed for each chromosome. Each point represent one locus from one conformation. The colors encode the ATAC-seq signal values.}
\label{fig:fig15}
\end{figure*}

Each monomer/locus in the 3D structures generated is assigned a value representing its ATAC signal. We use ATAC BED file from GEO repository GSE47753. The original data, however, needed to be processed in order to use in conjunction with our model. The procedure is illustrated in Fig.\ref{fig:atac-seq-illustration}. Each line in the BED file corresponds to a ATAC peak, associated with the peak value and the start and end genomic positions of the segment. In our model, each monomer represents a 100kbps genome segment. We count how many basepairs are overlapped between the segment represented by a single locus in our model and the segment in the ATAC-seq data. The contribution of the locus to ATAC signal value is computed proportionally from the peak value. For instance, the segment in the ATAC data that has a peak value of 100, and whose length is 50 kpbs, would have an overlap of length 30kbps with te locus. Then the contribution of ATAC signal from the segment in the ATAC data is $(30/50)*100=60$. If a segment has no data in the ATAC BED file, we set the peak value to zero.

\section*{Appendix J: Code  availability}
The code for the HIPPS method presented in this work and its detailed user instruction can be accessed at the Github repository \url{https://github.com/anyuzx/HIPPS-DIMES}.

The program is used as a Python script. The script accepts a Hi-C contact map or a mean spatial distance map as an input, and generates an ensemble of individual conformations. The Hi-C contact map can be in either \texttt{cooler} format or pure text format. The output conformations are in \texttt{.xyz} format, which users can use to compute various quantities of interest or can be rendered using VMD or other compatible softwares.

The script accepts a number of options. A partial list of available options are the following,

\begin{itemize}
	\item Number of individual conformations to be generated.
	\item Number of iterations of iterative scaling
	\item Value of learning rate $r$ in Eq.\ref{eq:30}
	\item The Chromosome region of interest
\end{itemize}

A detailed set of instructions and examples are provided on the Gihub page.

\begin{figure*}
\includegraphics[width=\linewidth]{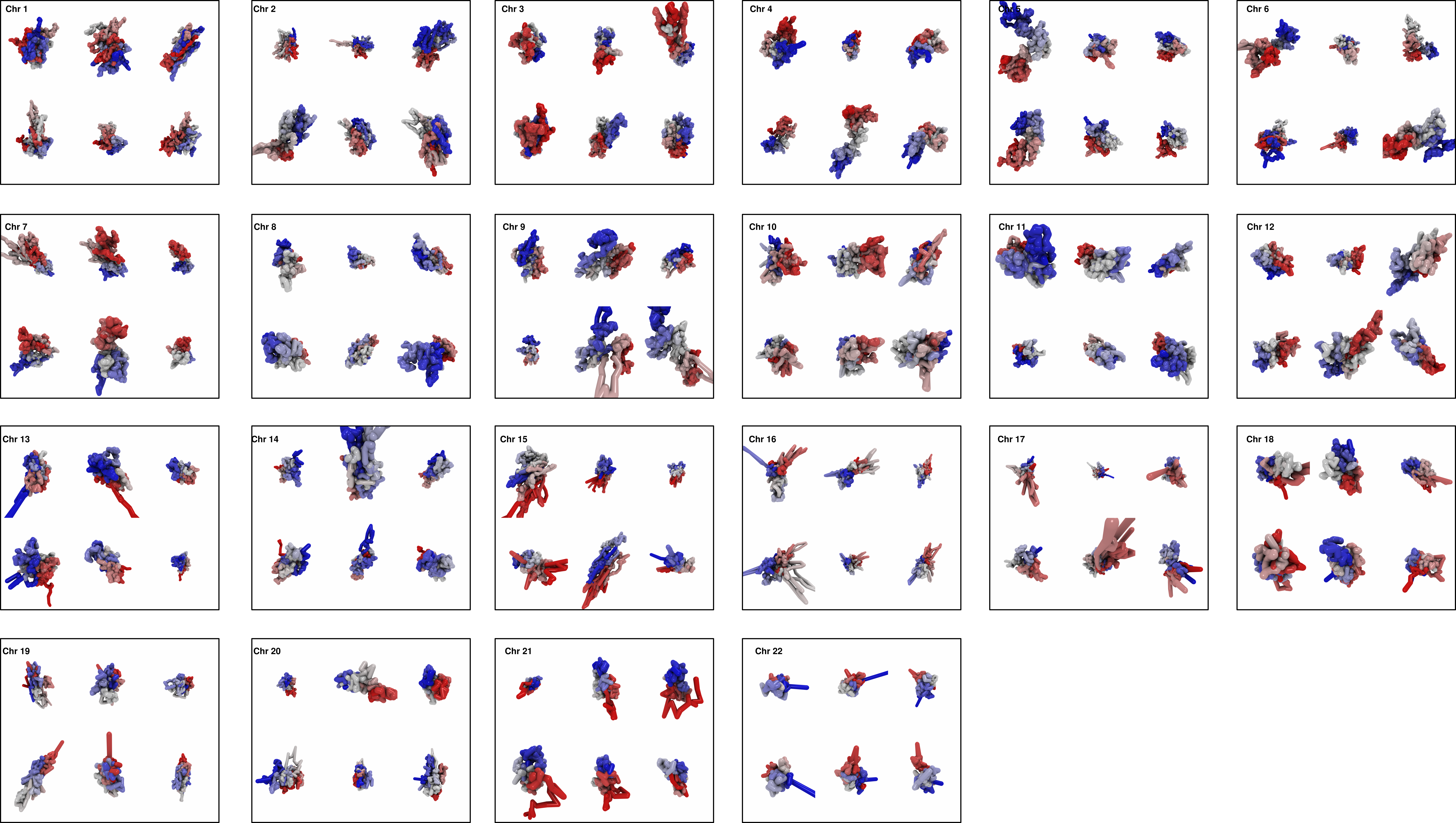}
\caption{More individual conformation for all 23 chromosomes for GM12878 cell line. Six individual conformations are shown for each chromosome. The colors encode the genomic position of the loci. Red and blue represent the 5' and 3' ends, respectively. The resolution of loci is 100 kbps. }
\label{fig:gm12878_individual_conformations}
\end{figure*}

\begin{figure*}
\includegraphics[width=\linewidth]{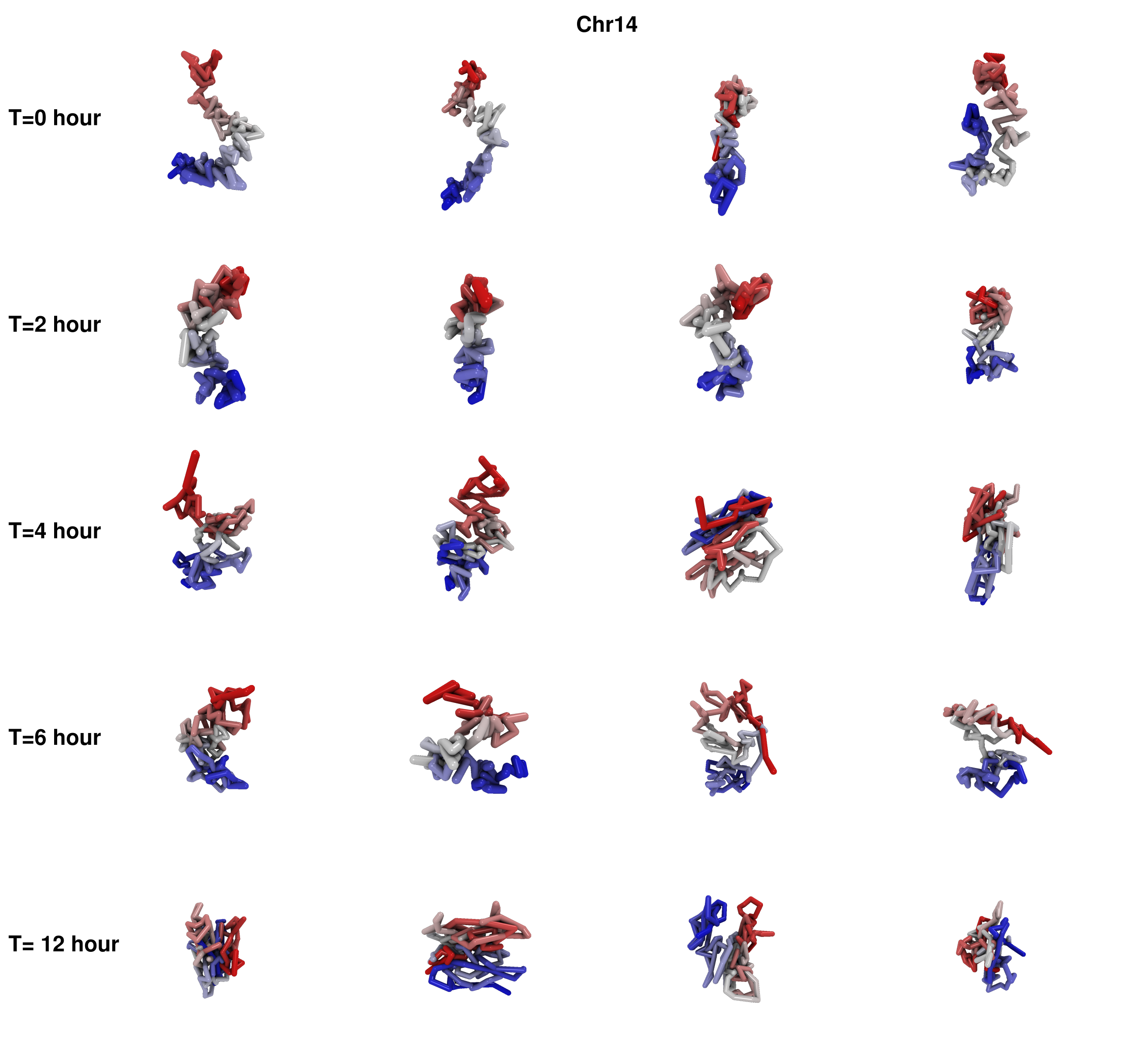}
\caption{Individual conformation for chromosome 14 for HeLa cell line at 0, 2, 4, 6, 12 hours. Four individual conformations are shown for each time point. Each individual conformation is generated randomly. The colors encode the genomic position of the loci. Red and blue represent the 5' and 3' ends, respectively. }
\label{fig:abramo2019_chr14_individual_structures}
\end{figure*}

\begin{figure*}
\includegraphics[width=\linewidth]{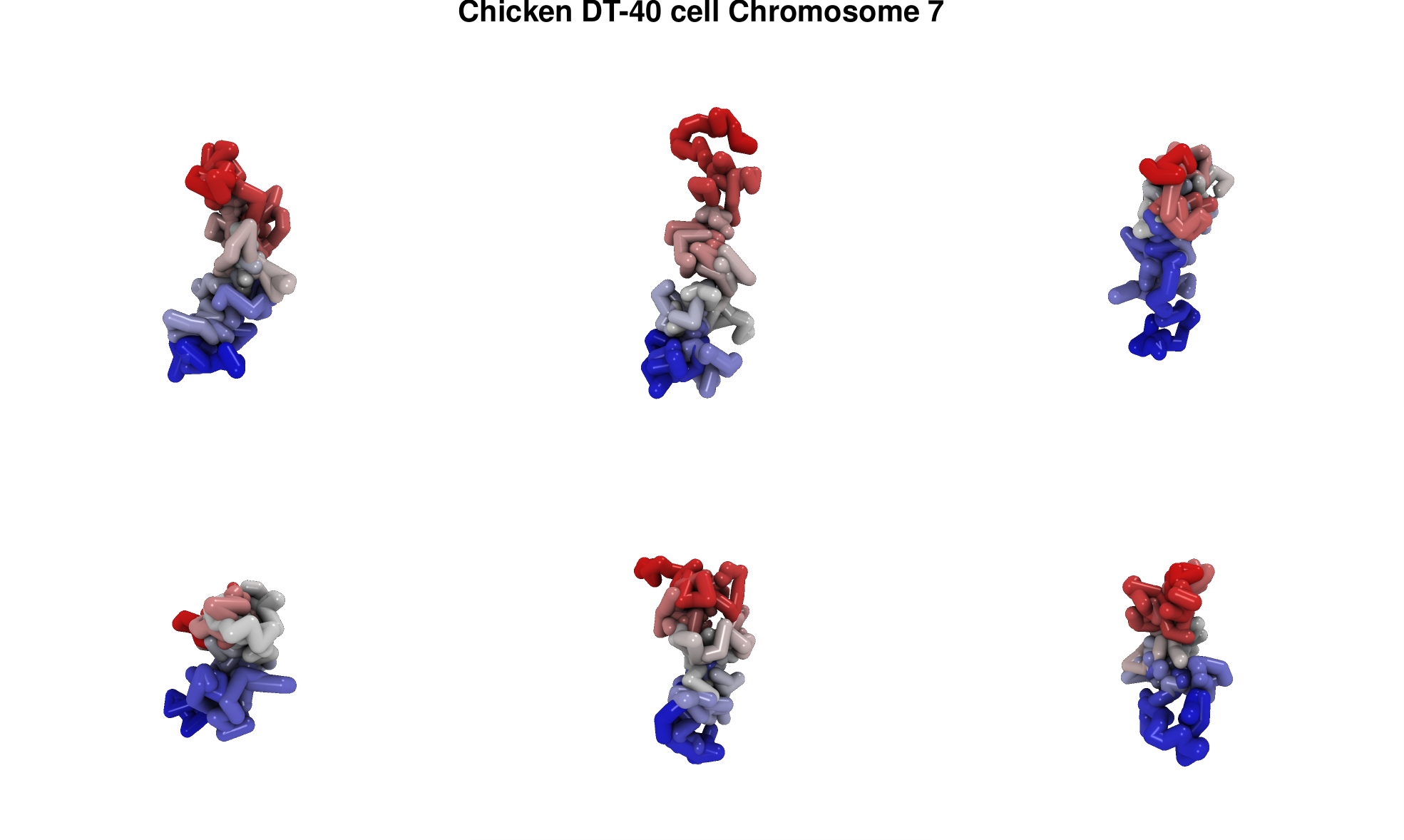}
\caption{Individual conformation for chromosome 7 for chicken DT-40 cell line. Six individual conformations are shown. Each individual conformation is generated randomly. The colors encode the genomic position of the loci. Red and blue represent the 5' and 3' ends, respectively. }
\label{fig:chicken_mitotic_alph30_individual_structure}
\end{figure*}

\clearpage

\bigskip
\textbf{Acknowledgements:} We are grateful to Atreya Dey and Sucheol Shin for using the Github code and providing important  feedback. We are grateful to the National Science Foundation (CHE 19-00093) and the Collie-Welch Regents Chair (F-0019) for supporting this work.


\begin{thebibliography}{10}
\expandafter\ifx\csname url\endcsname\relax
  \def\url#1{\texttt{#1}}\fi
\expandafter\ifx\csname urlprefix\endcsname\relax\def\urlprefix{URL }\fi
\providecommand{\bibinfo}[2]{#2}
\providecommand{\eprint}[2][]{\url{#2}}

\bibitem{LiebermanAiden2009}
\bibinfo{author}{Lieberman-Aiden, E.} \emph{et~al.}
\newblock \bibinfo{title}{Comprehensive mapping of long-range interactions
  reveals folding principles of the human genome}.
\newblock \emph{\bibinfo{journal}{Science}} \textbf{\bibinfo{volume}{326}},
  \bibinfo{pages}{289--293} (\bibinfo{year}{2009}).
\newblock \urlprefix\url{https://doi.org/10.1126/science.1181369}.

\bibitem{Dixon2012}
\bibinfo{author}{Dixon, J.~R.} \emph{et~al.}
\newblock \bibinfo{title}{Topological domains in mammalian genomes identified
  by analysis of chromatin interactions}.
\newblock \emph{\bibinfo{journal}{Nature}} \textbf{\bibinfo{volume}{485}},
  \bibinfo{pages}{376--380} (\bibinfo{year}{2012}).
\newblock \urlprefix\url{https://doi.org/10.1038/nature11082}.

\bibitem{Sexton2012}
\bibinfo{author}{Sexton, T.} \emph{et~al.}
\newblock \bibinfo{title}{Three-dimensional folding and functional organization
  principles of the drosophila genome}.
\newblock \emph{\bibinfo{journal}{Cell}} \textbf{\bibinfo{volume}{148}},
  \bibinfo{pages}{458--472} (\bibinfo{year}{2012}).
\newblock \urlprefix\url{https://doi.org/10.1016/j.cell.2012.01.010}.

\bibitem{Jin2013}
\bibinfo{author}{Jin, F.} \emph{et~al.}
\newblock \bibinfo{title}{A high-resolution map of the three-dimensional
  chromatin interactome in human cells}.
\newblock \emph{\bibinfo{journal}{Nature}} \textbf{\bibinfo{volume}{503}},
  \bibinfo{pages}{290--294} (\bibinfo{year}{2013}).
\newblock \urlprefix\url{https://doi.org/10.1038/nature12644}.

\bibitem{Dekker2013}
\bibinfo{author}{Dekker, J.}, \bibinfo{author}{Marti-Renom, M.~A.} \&
  \bibinfo{author}{Mirny, L.~A.}
\newblock \bibinfo{title}{Exploring the three-dimensional organization of
  genomes: interpreting chromatin interaction data}.
\newblock \emph{\bibinfo{journal}{Nature Reviews Genetics}}
  \textbf{\bibinfo{volume}{14}}, \bibinfo{pages}{390--403}
  (\bibinfo{year}{2013}).
\newblock \urlprefix\url{https://doi.org/10.1038/nrg3454}.

\bibitem{rao20143d}
\bibinfo{author}{Rao, S.~S.} \emph{et~al.}
\newblock \bibinfo{title}{{A 3D map of the human genome at kilobase resolution
  reveals principles of chromatin looping}}.
\newblock \emph{\bibinfo{journal}{Cell}} \textbf{\bibinfo{volume}{159}},
  \bibinfo{pages}{1665--1680} (\bibinfo{year}{2014}).

\bibitem{wang2016spatial}
\bibinfo{author}{Wang, S.} \emph{et~al.}
\newblock \bibinfo{title}{{Spatial organization of chromatin domains and
  compartments in single chromosomes}}.
\newblock \emph{\bibinfo{journal}{Science}} \textbf{\bibinfo{volume}{353}},
  \bibinfo{pages}{598--602} (\bibinfo{year}{2016}).

\bibitem{Shi2019}
\bibinfo{author}{Shi, G.} \& \bibinfo{author}{Thirumalai, D.}
\newblock \bibinfo{title}{Conformational heterogeneity in human interphase
  chromosome organization reconciles the {FISH} and hi-c paradox}.
\newblock \emph{\bibinfo{journal}{Nature Communications}}
  \textbf{\bibinfo{volume}{10}} (\bibinfo{year}{2019}).
\newblock \urlprefix\url{https://doi.org/10.1038/s41467-019-11897-0}.

\bibitem{bryngelson1996internal}
\bibinfo{author}{Bryngelson, J.} \& \bibinfo{author}{Thirumalai, D.}
\newblock \bibinfo{title}{{Internal constraints induce localization in an
  isolated polymer molecule}}.
\newblock \emph{\bibinfo{journal}{Phys. Rev. Lett.}}
  \textbf{\bibinfo{volume}{76}}, \bibinfo{pages}{542} (\bibinfo{year}{1996}).

\bibitem{Duan2010}
\bibinfo{author}{Duan, Z.} \emph{et~al.}
\newblock \bibinfo{title}{A three-dimensional model of the yeast genome}.
\newblock \emph{\bibinfo{journal}{Nature}} \textbf{\bibinfo{volume}{465}},
  \bibinfo{pages}{363--367} (\bibinfo{year}{2010}).
\newblock \urlprefix\url{https://doi.org/10.1038/nature08973}.

\bibitem{Kalhor2011}
\bibinfo{author}{Kalhor, R.}, \bibinfo{author}{Tjong, H.},
  \bibinfo{author}{Jayathilaka, N.}, \bibinfo{author}{Alber, F.} \&
  \bibinfo{author}{Chen, L.}
\newblock \bibinfo{title}{Genome architectures revealed by tethered chromosome
  conformation capture and population-based modeling}.
\newblock \emph{\bibinfo{journal}{Nature Biotechnology}}
  \textbf{\bibinfo{volume}{30}}, \bibinfo{pages}{90--98}
  (\bibinfo{year}{2011}).
\newblock \urlprefix\url{https://doi.org/10.1038/nbt.2057}.

\bibitem{Rousseau2011}
\bibinfo{author}{Rousseau, M.}, \bibinfo{author}{Fraser, J.},
  \bibinfo{author}{Ferraiuolo, M.~A.}, \bibinfo{author}{Dostie, J.} \&
  \bibinfo{author}{Blanchette, M.}
\newblock \bibinfo{title}{Three-dimensional modeling of chromatin structure
  from interaction frequency data using markov chain monte carlo sampling}.
\newblock \emph{\bibinfo{journal}{{BMC} Bioinformatics}}
  \textbf{\bibinfo{volume}{12}}, \bibinfo{pages}{414} (\bibinfo{year}{2011}).
\newblock \urlprefix\url{https://doi.org/10.1186/1471-2105-12-414}.

\bibitem{Zhang2013}
\bibinfo{author}{Zhang, Z.}, \bibinfo{author}{Li, G.}, \bibinfo{author}{Toh,
  K.-C.} \& \bibinfo{author}{Sung, W.-K.}
\newblock \bibinfo{title}{3d chromosome modeling with semi-definite programming
  and hi-c data}.
\newblock \emph{\bibinfo{journal}{Journal of Computational Biology}}
  \textbf{\bibinfo{volume}{20}}, \bibinfo{pages}{831--846}
  (\bibinfo{year}{2013}).
\newblock \urlprefix\url{https://doi.org/10.1089/cmb.2013.0076}.

\bibitem{Hu2013}
\bibinfo{author}{Hu, M.} \emph{et~al.}
\newblock \bibinfo{title}{Bayesian inference of spatial organizations of
  chromosomes}.
\newblock \emph{\bibinfo{journal}{{PLoS} Computational Biology}}
  \textbf{\bibinfo{volume}{9}}, \bibinfo{pages}{e1002893}
  (\bibinfo{year}{2013}).
\newblock \urlprefix\url{https://doi.org/10.1371/journal.pcbi.1002893}.

\bibitem{Varoquaux2014}
\bibinfo{author}{Varoquaux, N.}, \bibinfo{author}{Ay, F.},
  \bibinfo{author}{Noble, W.~S.} \& \bibinfo{author}{Vert, J.-P.}
\newblock \bibinfo{title}{A statistical approach for inferring the 3d structure
  of the genome}.
\newblock \emph{\bibinfo{journal}{Bioinformatics}}
  \textbf{\bibinfo{volume}{30}}, \bibinfo{pages}{i26--i33}
  (\bibinfo{year}{2014}).
\newblock \urlprefix\url{https://doi.org/10.1093/bioinformatics/btu268}.

\bibitem{Lesne2014}
\bibinfo{author}{Lesne, A.}, \bibinfo{author}{Riposo, J.},
  \bibinfo{author}{Roger, P.}, \bibinfo{author}{Cournac, A.} \&
  \bibinfo{author}{Mozziconacci, J.}
\newblock \bibinfo{title}{3d genome reconstruction from chromosomal contacts}.
\newblock \emph{\bibinfo{journal}{Nature Methods}}
  \textbf{\bibinfo{volume}{11}}, \bibinfo{pages}{1141--1143}
  (\bibinfo{year}{2014}).
\newblock \urlprefix\url{https://doi.org/10.1038/nmeth.3104}.

\bibitem{Tjong2016}
\bibinfo{author}{Tjong, H.} \emph{et~al.}
\newblock \bibinfo{title}{Population-based 3d genome structure analysis reveals
  driving forces in spatial genome organization}.
\newblock \emph{\bibinfo{journal}{Proceedings of the National Academy of
  Sciences}} \textbf{\bibinfo{volume}{113}}, \bibinfo{pages}{E1663--E1672}
  (\bibinfo{year}{2016}).
\newblock \urlprefix\url{https://doi.org/10.1073/pnas.1512577113}.

\bibitem{Hua2018}
\bibinfo{author}{Hua, N.} \emph{et~al.}
\newblock \bibinfo{title}{Producing genome structure populations with the
  dynamic and automated {PGS} software}.
\newblock \emph{\bibinfo{journal}{Nature Protocols}}
  \textbf{\bibinfo{volume}{13}}, \bibinfo{pages}{915--926}
  (\bibinfo{year}{2018}).
\newblock \urlprefix\url{https://doi.org/10.1038/nprot.2018.008}.

\bibitem{Giorgetti2014}
\bibinfo{author}{Giorgetti, L.} \emph{et~al.}
\newblock \bibinfo{title}{Predictive polymer modeling reveals coupled
  fluctuations in chromosome conformation and transcription}.
\newblock \emph{\bibinfo{journal}{Cell}} \textbf{\bibinfo{volume}{157}},
  \bibinfo{pages}{950--963} (\bibinfo{year}{2014}).
\newblock \urlprefix\url{https://doi.org/10.1016/j.cell.2014.03.025}.

\bibitem{Zhang2015}
\bibinfo{author}{Zhang, B.} \& \bibinfo{author}{Wolynes, P.~G.}
\newblock \bibinfo{title}{Topology, structures, and energy landscapes of human
  chromosomes}.
\newblock \emph{\bibinfo{journal}{Proceedings of the National Academy of
  Sciences}} \textbf{\bibinfo{volume}{112}}, \bibinfo{pages}{6062--6067}
  (\bibinfo{year}{2015}).
\newblock \urlprefix\url{https://doi.org/10.1073/pnas.1506257112}.

\bibitem{LeTreut2018}
\bibinfo{author}{Treut, G.~L.}, \bibinfo{author}{K{\'{e}}p{\`{e}}s, F.} \&
  \bibinfo{author}{Orland, H.}
\newblock \bibinfo{title}{A polymer model for the quantitative reconstruction
  of chromosome architecture from {HiC} and {GAM} data}.
\newblock \emph{\bibinfo{journal}{Biophysical Journal}}
  \textbf{\bibinfo{volume}{115}}, \bibinfo{pages}{2286--2294}
  (\bibinfo{year}{2018}).
\newblock \urlprefix\url{https://doi.org/10.1016/j.bpj.2018.10.032}.

\bibitem{Giorgetti2016}
\bibinfo{author}{Giorgetti, L.} \& \bibinfo{author}{Heard, E.}
\newblock \bibinfo{title}{Closing the loop: 3c versus {DNA} {FISH}}.
\newblock \emph{\bibinfo{journal}{Genome Biology}}
  \textbf{\bibinfo{volume}{17}} (\bibinfo{year}{2016}).
\newblock \urlprefix\url{https://doi.org/10.1186/s13059-016-1081-2}.

\bibitem{fudenberg2017fish}
\bibinfo{author}{Fudenberg, G.} \& \bibinfo{author}{Imakaev, M.}
\newblock \bibinfo{title}{{FISH-ing for captured contacts: towards reconciling
  FISH and 3C}}.
\newblock \emph{\bibinfo{journal}{Nature Methods}}  (\bibinfo{year}{2017}).

\bibitem{Bickmore2013}
\bibinfo{author}{Bickmore, W.~A.} \& \bibinfo{author}{van Steensel, B.}
\newblock \bibinfo{title}{{Genome Architecture: Domain Organization of
  Interphase Chromosomes}}.
\newblock \emph{\bibinfo{journal}{Cell}} \textbf{\bibinfo{volume}{152}},
  \bibinfo{pages}{1270--1284} (\bibinfo{year}{2013}).
\newblock \urlprefix\url{https://doi.org/10.1016/j.cell.2013.02.001}.

\bibitem{williamson2014spatial}
\bibinfo{author}{Williamson, I.} \emph{et~al.}
\newblock \bibinfo{title}{{Spatial genome organization: contrasting views from
  chromosome conformation capture and fluorescence in situ hybridization}}.
\newblock \emph{\bibinfo{journal}{Gene. Dev.}} \textbf{\bibinfo{volume}{28}},
  \bibinfo{pages}{2778--2791} (\bibinfo{year}{2014}).

\bibitem{Finn2019}
\bibinfo{author}{Finn, E.~H.} \emph{et~al.}
\newblock \bibinfo{title}{Extensive heterogeneity and intrinsic variation in
  spatial genome organization}.
\newblock \emph{\bibinfo{journal}{Cell}} \textbf{\bibinfo{volume}{176}},
  \bibinfo{pages}{1502--1515.e10} (\bibinfo{year}{2019}).
\newblock \urlprefix\url{https://doi.org/10.1016/j.cell.2019.01.020}.

\bibitem{Stevens2017}
\bibinfo{author}{Stevens, T.~J.} \emph{et~al.}
\newblock \bibinfo{title}{{3D structures of individual mammalian genomes
  studied by single-cell Hi-C}}.
\newblock \emph{\bibinfo{journal}{Nature}} \textbf{\bibinfo{volume}{544}},
  \bibinfo{pages}{59--64} (\bibinfo{year}{2017}).
\newblock \urlprefix\url{https://doi.org/10.1038/nature21429}.

\bibitem{lee2017topological}
\bibinfo{author}{Lee, H.}, \bibinfo{author}{Ma, Z.}, \bibinfo{author}{Wang, Y.}
  \& \bibinfo{author}{Chung, M.~K.}
\newblock \bibinfo{title}{{Topological Distances between Networks and Its
  Application to Brain Imaging}}.
\newblock \emph{\bibinfo{journal}{arXiv preprint arXiv:1701.04171}}
  (\bibinfo{year}{2017}).

\bibitem{Shi2018}
\bibinfo{author}{Shi, G.}, \bibinfo{author}{Liu, L.}, \bibinfo{author}{Hyeon,
  C.} \& \bibinfo{author}{Thirumalai, D.}
\newblock \bibinfo{title}{Interphase human chromosome exhibits out of
  equilibrium glassy dynamics}.
\newblock \emph{\bibinfo{journal}{Nature Communications}}
  \textbf{\bibinfo{volume}{9}} (\bibinfo{year}{2018}).
\newblock \urlprefix\url{https://doi.org/10.1038/s41467-018-05606-6}.

\bibitem{Fudenberg2017}
\bibinfo{author}{Fudenberg, G.} \& \bibinfo{author}{Imakaev, M.}
\newblock \bibinfo{title}{{FISH}-ing for captured contacts: towards reconciling
  {FISH} and 3c}.
\newblock \emph{\bibinfo{journal}{Nature Methods}}
  \textbf{\bibinfo{volume}{14}}, \bibinfo{pages}{673--678}
  (\bibinfo{year}{2017}).
\newblock \urlprefix\url{https://doi.org/10.1038/nmeth.4329}.

\bibitem{branco2006intermingling}
\bibinfo{author}{Branco, M.~R.} \& \bibinfo{author}{Pombo, A.}
\newblock \bibinfo{title}{{Intermingling of chromosome territories in
  interphase suggests role in translocations and transcription-dependent
  associations}}.
\newblock \emph{\bibinfo{journal}{PLoS Biol.}} \textbf{\bibinfo{volume}{4}},
  \bibinfo{pages}{e138} (\bibinfo{year}{2006}).

\bibitem{Rosa2008}
\bibinfo{author}{Rosa, A.} \& \bibinfo{author}{Everaers, R.}
\newblock \bibinfo{title}{Structure and dynamics of interphase chromosomes}.
\newblock \emph{\bibinfo{journal}{{PLoS} Computational Biology}}
  \textbf{\bibinfo{volume}{4}}, \bibinfo{pages}{e1000153}
  (\bibinfo{year}{2008}).
\newblock \urlprefix\url{https://doi.org/10.1371/journal.pcbi.1000153}.

\bibitem{DiPierro2016}
\bibinfo{author}{Pierro, M.~D.}, \bibinfo{author}{Zhang, B.},
  \bibinfo{author}{Aiden, E.~L.}, \bibinfo{author}{Wolynes, P.~G.} \&
  \bibinfo{author}{Onuchic, J.~N.}
\newblock \bibinfo{title}{Transferable model for chromosome architecture}.
\newblock \emph{\bibinfo{journal}{Proceedings of the National Academy of
  Sciences}} \textbf{\bibinfo{volume}{113}}, \bibinfo{pages}{12168--12173}
  (\bibinfo{year}{2016}).
\newblock \urlprefix\url{https://doi.org/10.1073/pnas.1613607113}.

\bibitem{Farr2018}
\bibinfo{author}{Farr{\'{e}}, P.} \& \bibinfo{author}{Emberly, E.}
\newblock \bibinfo{title}{A maximum-entropy model for predicting chromatin
  contacts}.
\newblock \emph{\bibinfo{journal}{{PLOS} Computational Biology}}
  \textbf{\bibinfo{volume}{14}}, \bibinfo{pages}{e1005956}
  (\bibinfo{year}{2018}).
\newblock \urlprefix\url{https://doi.org/10.1371/journal.pcbi.1005956}.

\bibitem{Liu2019}
\bibinfo{author}{Liu, L.} \& \bibinfo{author}{Kim, M. H.} \& \bibinfo{author}{Hyeon, C}
\newblock \bibinfo{title}{Heterogeneous Loop Model to Infer 3D Chromosome Structures from Hi-C}.
\newblock \emph{\bibinfo{journal}{Biophysical Journal}}
  \textbf{\bibinfo{volume}{117}}, \bibinfo{pages}{613--625}
  (\bibinfo{year}{2019}).
  
 
\bibitem{Darroch1972}
\bibinfo{author}{Darroch, J.~N.} \& \bibinfo{author}{Ratcliff, D.}
\newblock \bibinfo{title}{Generalized iterative scaling for log-linear models}.
\newblock \emph{\bibinfo{journal}{The Annals of Mathematical Statistics}}
  \textbf{\bibinfo{volume}{43}}, \bibinfo{pages}{1470--1480}
  (\bibinfo{year}{1972}).
\newblock \urlprefix\url{http://www.jstor.org/stable/2240069}.

\bibitem{berger1997improved}
\bibinfo{author}{Berger, A.}
\newblock \bibinfo{title}{The improved iterative scaling algorithm: A gentle
  introduction} (\bibinfo{year}{1997}).

\bibitem{grosberg1988role}
\bibinfo{author}{Grosberg, A.~Y.}, \bibinfo{author}{Nechaev, S.~K.} \&
  \bibinfo{author}{Shakhnovich, E.~I.}
\newblock \bibinfo{title}{{The role of topological constraints in the kinetics
  of collapse of macromolecules}}.
\newblock \emph{\bibinfo{journal}{J. Phys-paris.}}
  \textbf{\bibinfo{volume}{49}}, \bibinfo{pages}{2095--2100}
  (\bibinfo{year}{1988}).

\bibitem{grosberg1993crumpled}
\bibinfo{author}{Grosberg, A.}, \bibinfo{author}{Rabin, Y.},
  \bibinfo{author}{Havlin, S.} \& \bibinfo{author}{Neer, A.}
\newblock \bibinfo{title}{{Crumpled globule model of the three-dimensional
  structure of DNA}}.
\newblock \emph{\bibinfo{journal}{Europhys. Lett.}}
  \textbf{\bibinfo{volume}{23}}, \bibinfo{pages}{373} (\bibinfo{year}{1993}).

\bibitem{lieberman2009comprehensive}
\bibinfo{author}{Lieberman-Aiden, E.} \emph{et~al.}
\newblock \bibinfo{title}{{Comprehensive mapping of long-range interactions
  reveals folding principles of the human genome}}.
\newblock \emph{\bibinfo{journal}{Science}} \textbf{\bibinfo{volume}{326}},
  \bibinfo{pages}{289--293} (\bibinfo{year}{2009}).

\bibitem{Buenrostro2013}
\bibinfo{author}{Buenrostro, J.~D.}, \bibinfo{author}{Giresi, P.~G.},
  \bibinfo{author}{Zaba, L.~C.}, \bibinfo{author}{Chang, H.~Y.} \&
  \bibinfo{author}{Greenleaf, W.~J.}
\newblock \bibinfo{title}{Transposition of native chromatin for fast and
  sensitive epigenomic profiling of open chromatin, {DNA}-binding proteins and
  nucleosome position}.
\newblock \emph{\bibinfo{journal}{Nature Methods}}
  \textbf{\bibinfo{volume}{10}}, \bibinfo{pages}{1213--1218}
  (\bibinfo{year}{2013}).
\newblock \urlprefix\url{https://doi.org/10.1038/nmeth.2688}.

\bibitem{maaten2008visualizing}
\bibinfo{author}{Maaten, L. v.~d.} \& \bibinfo{author}{Hinton, G.}
\newblock \bibinfo{title}{{Visualizing data using t-SNE}}.
\newblock \emph{\bibinfo{journal}{J. Mach. Learn. Res.}}
  \textbf{\bibinfo{volume}{9}}, \bibinfo{pages}{2579--2605}
  (\bibinfo{year}{2008}).

\bibitem{Abramo2019}
\bibinfo{author}{Abramo, K.} \emph{et~al.}
\newblock \bibinfo{title}{A chromosome folding intermediate at the
  condensin-to-cohesin transition during telophase}.
\newblock \emph{\bibinfo{journal}{Nature Cell Biology}}
  \textbf{\bibinfo{volume}{21}}, \bibinfo{pages}{1393--1402}
  (\bibinfo{year}{2019}).
\newblock \urlprefix\url{https://doi.org/10.1038/s41556-019-0406-2}.

\bibitem{Gibcus2018}
\bibinfo{author}{Gibcus, J.~H.} \emph{et~al.}
\newblock \bibinfo{title}{A pathway for mitotic chromosome formation}.
\newblock \emph{\bibinfo{journal}{Science}} \textbf{\bibinfo{volume}{359}},
  \bibinfo{pages}{eaao6135} (\bibinfo{year}{2018}).
\newblock \urlprefix\url{https://doi.org/10.1126/science.aao6135}.

\bibitem{honeycutt1992nature}
\bibinfo{author}{Honeycutt, J.} \& \bibinfo{author}{Thirumalai, D.}
\newblock \bibinfo{title}{{The nature of folded states of globular proteins}}.
\newblock \emph{\bibinfo{journal}{Biopolymers}} \textbf{\bibinfo{volume}{32}},
  \bibinfo{pages}{695--709} (\bibinfo{year}{1992}).

\bibitem{des1980short}
\bibinfo{author}{Des~Cloizeaux, J.}
\newblock \bibinfo{title}{Short range correlation between elements of a long
  polymer in a good solvent}.
\newblock \emph{\bibinfo{journal}{Journal de Physique}}
  \textbf{\bibinfo{volume}{41}}, \bibinfo{pages}{223--238}
  (\bibinfo{year}{1980}).

\bibitem{bohn2009conformational}
\bibinfo{author}{Bohn, M.} \& \bibinfo{author}{Heermann, D.~W.}
\newblock \bibinfo{title}{Conformational properties of compact polymers}.
\newblock \emph{\bibinfo{journal}{The Journal of chemical physics}}
  \textbf{\bibinfo{volume}{130}}, \bibinfo{pages}{174901}
  (\bibinfo{year}{2009}).
 
\end{thebibliography}
\end{document}